\documentclass[preprint]{aastex}
\usepackage[usenames,dvipsnames]{color}
\usepackage{natbib,graphicx,latexsym,lscape,pdflscape,url,pdfpages}

\newcommand{\Gaia}{{\sl Gaia}}

\newcommand{\WISE}{{\sl WISE}}

\newcommand{\Msun}{\mbox{$M_{\sun}$}}

\newcommand{\Lsun}{\mbox{$L_{\sun}$}}

\newcommand{\Mjup}{\mbox{$M_{\rm Jup}$}}

\newcommand{\perpix}{\mbox{pixel$^{-1}$}}

\newcommand{\kms}{\mbox{km\,s$^{-1}$}}
\newcommand{\masyr}{\hbox{mas\,yr$^{-1}$}}

\newcommand{\Ks}{\mbox{$K_S$}}

\newcommand{\Kn}{\mbox{$K_{\rm H2}$}}

\newcommand{\Lbol}{\mbox{$L_{\rm bol}$}}
\newcommand{\fbol}{\mbox{$f_{\rm bol}$}}

\newcommand{\Mbol}{\mbox{$M_{\rm bol}$}}
\newcommand{\mbol}{\mbox{$m_{\rm bol}$}}

\newcommand{\vsini}{\mbox{$v\sin(i)$}}

\newcommand{\vlg}{\textsc{vl-g}}
\newcommand{\intg}{\textsc{int-g}}
\newcommand{\fldg}{\textsc{fld-g}}

\newcommand{\objlong}{2MASS~J02495639$-$0557352}
\newcommand{\objlongb}{2MASS~J02495436$-$0558015}

\newcommand{\obj}{2MASS~J0249$-$0557}
\newcommand{\objtwo}{2MASS~J2208+2921} 
\newcommand{\bpic}{$\beta$~Pic}

\shorttitle{\obj~c}
\shortauthors{Dupuy et al.}

\begin{document}

\title{The Hawaii Infrared Parallax Program.  III.
  2MASS~J0249$-$0557~c: A Wide Planetary-mass Companion to a Low-Mass
  Binary in the $\beta$~Pic Moving Group\altaffilmark{*\dag}}

\author{Trent J.\ Dupuy,\altaffilmark{1} 
        Michael C.\ Liu,\altaffilmark{2}
        Katelyn N.\ Allers,\altaffilmark{3,4} 
        Beth A.\ Biller,\altaffilmark{4,5,6} 
        Kaitlin M.\ Kratter,\altaffilmark{7} 
        Andrew W.\ Mann,\altaffilmark{8,9} 
        Evgenya L.\ Shkolnik,\altaffilmark{10}
        Adam L.\ Kraus,\altaffilmark{11}
        and William M.\ J.\ Best\altaffilmark{2}}

      \altaffiltext{*}{Based on data obtained with WIRCam, a joint
        project of CFHT, Taiwan, Korea, Canada, France, at the
        Canada-France-Hawaii Telescope, which is operated by the
        National Research Council of Canada, the Institute National
        des Sciences de l'Univers of the Centre National de la
        Recherche Scientifique of France, and the University of
        Hawaii.}

      \altaffiltext{\dag}{Data presented herein were obtained at the
        W.M.\ Keck Observatory, which is operated as a scientific
        partnership among the California Institute of Technology, the
        University of California, and the National Aeronautics and
        Space Administration. The Observatory was made possible by the
        generous financial support of the W.M.\ Keck Foundation.}

      \altaffiltext{1}{Gemini Observatory, Northern Operations Center,
        670 N.\ A'ohoku Place, Hilo, HI 96720, USA}

      \altaffiltext{2}{Institute for Astronomy, University of Hawaii,
        2680 Woodlawn Drive, Honolulu, HI 96822, USA}

      \altaffiltext{3}{Department of Physics and Astronomy, Bucknell
        University, Lewisburg, PA 17837, USA}

      \altaffiltext{4}{Visiting Astronomer at the Infrared Telescope
        Facility, which is operated by the University of Hawaii under
        contract NNH14CK55B with the National Aeronautics and Space
        Administration}

      \altaffiltext{5}{SUPA Institute for Astronomy, University of
        Edinburgh, Blackford Hill View, Edinburgh EH9 3HJ, UK}

      \altaffiltext{6}{Centre for Exoplanet Science, University of
        Edinburgh, UK}

      \altaffiltext{7}{Department of Astronomy, University of Arizona,
        933 N Cherry Ave, Tucson, AZ 85721, USA}

      \altaffiltext{8}{Department of Astronomy, Columbia University,
        550 West 120th Street, New York, NY 10027, USA}

      \altaffiltext{9}{NASA Hubble Fellow}

      \altaffiltext{10}{School of Earth and Space Exploration, Arizona
        State University, Tempe, AZ 85281, USA}

      \altaffiltext{11}{The University of Texas at Austin, Department
        of Astronomy, 2515 Speedway C1400, Austin, TX 78712, USA}

\begin{abstract}

  We have discovered a wide planetary-mass companion to the \bpic\
  moving group member \objlong\ (M6~\vlg) using CFHT/WIRCam astrometry
  from the Hawaii Infrared Parallax Program.  In addition, Keck laser
  guide star adaptive optics aperture-masking interferometry shows
  that the host is itself a tight binary.  Altogether, \obj{AB}c is a
  bound triple system with an $11.6^{+1.3}_{-1.0}$\,\Mjup\ object
  separated by $1950\pm200$\,AU (40\arcsec) from a relatively close
  ($2.17\pm0.22$\,AU, 0\farcs04) pair of $48^{+13}_{-12}$\,\Mjup\ and
  $44^{+14}_{-11}$\,\Mjup\ objects.  \obj{AB} is one of the few
  ultracool binaries to be discovered in a young moving group and the
  first confirmed in the \bpic\ moving group ($22\pm6$\,Myr).  The
  mass, absolute magnitudes, and spectral type of \obj~c (L2~\vlg) are
  remarkably similar to those of the planet \bpic~b (L2,
  $13.0^{+0.4}_{-0.3}$\,\Mjup).  We also find that the free-floating
  object 2MASS~J2208+2921 (L3~\vlg) is another possible \bpic\ moving
  group member with colors and absolute magnitudes similar to \bpic~b
  and \obj~c.  \bpic~b is the first directly imaged planet to have a
  ``twin,'' namely an object of comparable properties in the same
  stellar association.  Such directly imaged objects provide a unique
  opportunity to measure atmospheric composition, variability, and
  rotation across different pathways of assembling planetary-mass
  objects from the same natal material.

\end{abstract}



\section{Introduction}

The formation of gas giants is a critical phase in the assembly of
planetary systems from circumstellar disks.  Direct imaging is a key
method for studying such planets as it provides direct access to their
photospheres, which can be used to probe many physical properties
(e.g., composition, surface temperature, chemistry).  Because direct
imaging is intrinsically more sensitive to planets farther from their
host stars, many planetary-mass companions have been discovered at
wide separations ($\gtrsim$100\,AU) where it is not clear if they
could have arisen from disks \citep[e.g., see the review
of][]{2016PASP..128j2001B}.
In practice, this population of wide-separation companions provides an
opportunity to delineate possible formation pathways, since if they
formed differently than close-in gas giants, there may be evidence in
their orbits or spectra (e.g., elemental abundances;
\citealp{2010Icar..207..503H}).
The widest companions ($\gtrsim$10$^3$~AU; e.g.,
\citealp{2011ApJ...730L...9L}, \citealp{2014ApJ...787....5N},
\citealp{2016MNRAS.457.3191D}) offer the sharpest contrast with
directly imaged planets that are on close orbits, such as 51~Eri~b
\citep[13\,AU;][]{2015Sci...350...64M} and the HR~8799 system
\citep[14--68\,AU;][]{2008Sci...322.1348M, 2010Natur.468.1080M}.
Studying these two populations along with a third group, free-floating
planetary-mass objects, like PSO~J318.5338$-$22.8603
\citep{2013ApJ...777L..20L} and SDSS~J1110+0116
\citep{2015ApJ...808L..20G}, should offer a clearer picture of gas
giant formation.

Directly imaged planets that are members of stellar associations are
particularly valuable because the age and the composition of their
natal material can be constrained by the entire ensemble of stars in
the group.  There are relatively few close-in ($<$100\,AU) imaged
planets that have ages determined by being a member of a moving group
or association.
\bpic~b \citep[9~AU;][]{2010Sci...329...57L} and 51~Eri~b
\citep[13~AU;][]{2015Sci...350...64M} are members of the \bpic\ moving
group ($22\pm6$\,Myr; \citealp{2017AJ....154...69S}). 
HD~95086~b \citep[56~AU;][]{2013ApJ...772L..15R} and HIP~65426~b
\citep[82~AU;][]{2017A&A...605L...9C} are members of Lower Centaurus
Crux ($17\pm2$\,Myr; \citealp{2012ApJ...746..154P}).
2MASS~J1207$-$3932~b \citep[41~AU;][]{2004A&A...425L..29C} is a member
of the TW~Hydra association ($10\pm3$\,Myr;
\citealp{2015MNRAS.454..593B}).
LkCa~15 is a young Taurus member ($2^{+2}_{-1}$\,Myr;
\citealp{2009ApJ...704..531K}) that may host one or more planets
\citep[15--20~AU;][]{2012ApJ...745....5K, 2015Natur.527..342S}.
And the HR~8799 system \citep[14--68~AU;][]{2008Sci...322.1348M,
  2010Natur.468.1080M} is a proposed member of Columba
($42^{+6}_{-4}$\,Myr; \citealp{2011ApJ...732...61Z,
  2015MNRAS.454..593B}).
A few more $<$100\,AU companions have higher mass estimates that place
them near or above the deuterium-fusing limit: 2MASS~J0122$-$2439~B
\citep[52~AU;][]{2013ApJ...774...55B} is a possible member of
AB~Doradus ($150^{+50}_{-20}$\,Myr; \citealp{2015MNRAS.454..593B}),
while 2MASS~J0103$-$5515~b \citep[84~AU;][]{2013A&A...553L...5D} is a
member of Tucana-Horologium
\citep[$45\pm4$\,Myr;][]{2014AJ....147..146K, 2015MNRAS.454..593B}.

We present here a new planetary-mass companion in the \bpic\ moving
group discovered in seeing-limited astrometry from the Hawaii Infrared
Parallax Program \citep{2012ApJS..201...19D, 2016ApJ...833...96L}. In
addition, we have discovered that its host \objlong\ (hereinafter
\obj) is actually a tight, nearly equal-flux binary using
aperture-masking data obtained with Keck laser guide star adaptive
optics (LGS AO). The host \obj{AB} was originally identified (in
integrated light) as a member of the \bpic\ moving group by
\citet{2017AJ....154...69S} from its proper motion and radial velocity
(RV). In the following, we reaffirm this system's membership in the
\bpic\ moving group with a parallax and new proper motion and show
that the companion \obj~c is a physically bound object, making this a
triple system of very low-mass objects. We discuss this unique system
in the context of other \bpic\ members and other planetary-mass
companions.


\section{Observations}

\subsection{CFHT/WIRCam \label{sec:cfht}}

\begin{figure} 
  \vskip -0.3in
  \centerline{\includegraphics[width=6.5in,angle=0]{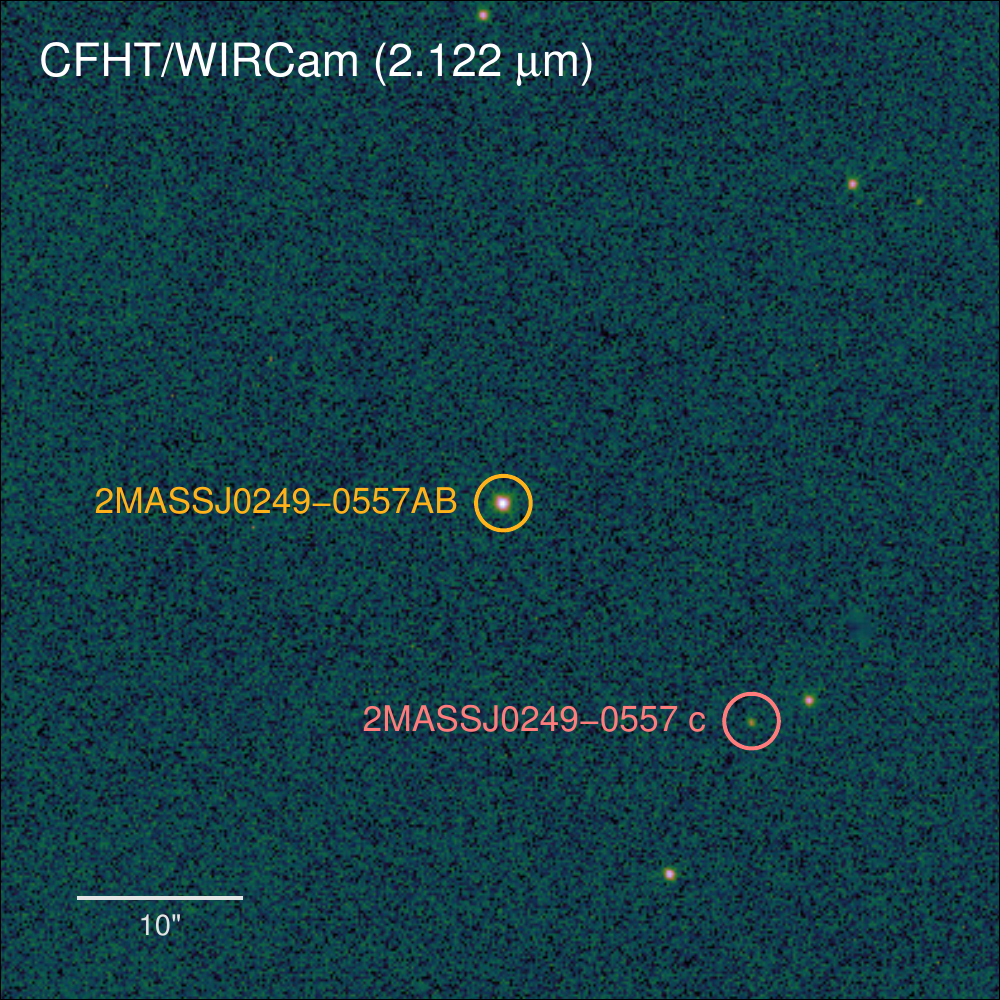}}
  \caption{\normalsize A $120\arcsec\times120\arcsec$ cutout from a
    single CFHT/WIRCam \Kn-band image ($t_{\rm exp}=5$\,s) typical of
    those used in our astrometric analysis.  This image was taken on
    2012~Aug~12~UT in 0\farcs62 seeing and is shown at its native
    orientation, within 0\fdg1 of north up and east left, using an
    asinh stretch. The image is centered on the target of our parallax
    observations (\obj{AB}, M6~\vlg), and the newly discovered
    companion is circled to the lower right. Five other unassociated
    reference stars are visible throughout this image, two of which
    are closer to \obj~c than its host star. \label{fig:image}}
\end{figure}

As part of our ongoing Hawaii Infrared Parallax Program at the
Canada-France-Hawaii Telescope (CFHT), we have been using the facility
infrared camera WIRCam \citep{2004SPIE.5492..978P} to monitor \obj\ in
order to confirm the \bpic\ membership of the latest-type objects
identified by \citet{2017AJ....154...69S}.
Because our observations were designed to measure the parallax of this
relatively bright M6 dwarf, we used a narrow-band filter
(0.032\,\micron\ bandwidth) in the $K$ band.  We refer to this filter
as \Kn\ band because it is centered at 2.122\,\micron, the wavelength
of the H$_2$~1-0~S(1) line.  Figure~\ref{fig:image} shows a portion of
one of our WIRCam images.

Our observing strategy and reduction pipeline are described in detail
in our previous work \citep{2012ApJS..201...19D, 2016ApJ...833...96L}.
Briefly, we measure relative astrometry of all stars above a threshold
signal-to-noise ratio (${\rm SNR} > 5$ for this analysis), first using
SExtractor \citep{1996A&AS..117..393B} to compute $(x,y)$ positions
and fluxes and then our custom pipeline for the following steps. The
astrometric uncertainty for a given object at a given epoch is the
standard error on the mean, computed using the internal astrometric
scatter across the dithered images at a single epoch. The accuracy of
these error estimates is later verified by examining the $\chi^2$ of
our final five-parameter parallax and proper motion fits to our
relative astrometry. The absolute calibration of our astrometry (e.g.,
the pixel scale) is determined by matching low-proper-motion sources
($<30$\,\masyr) that also appear in an external reference catalog. In
this case, all 24~of our low-proper-motion reference stars were in DR12
of the Sloan Digital Sky Survey \citep[SDSS;][]{2015ApJS..219...12A},
and the rms of our astrometry compared to SDSS after performing a
linear transformation was $0\farcs031$, which we expect is dominated
by the uncertainty in the SDSS relative astrometry.

\begin{figure} 
  \vskip -0.3in
  \centerline{\includegraphics[width=6.5in,angle=0]{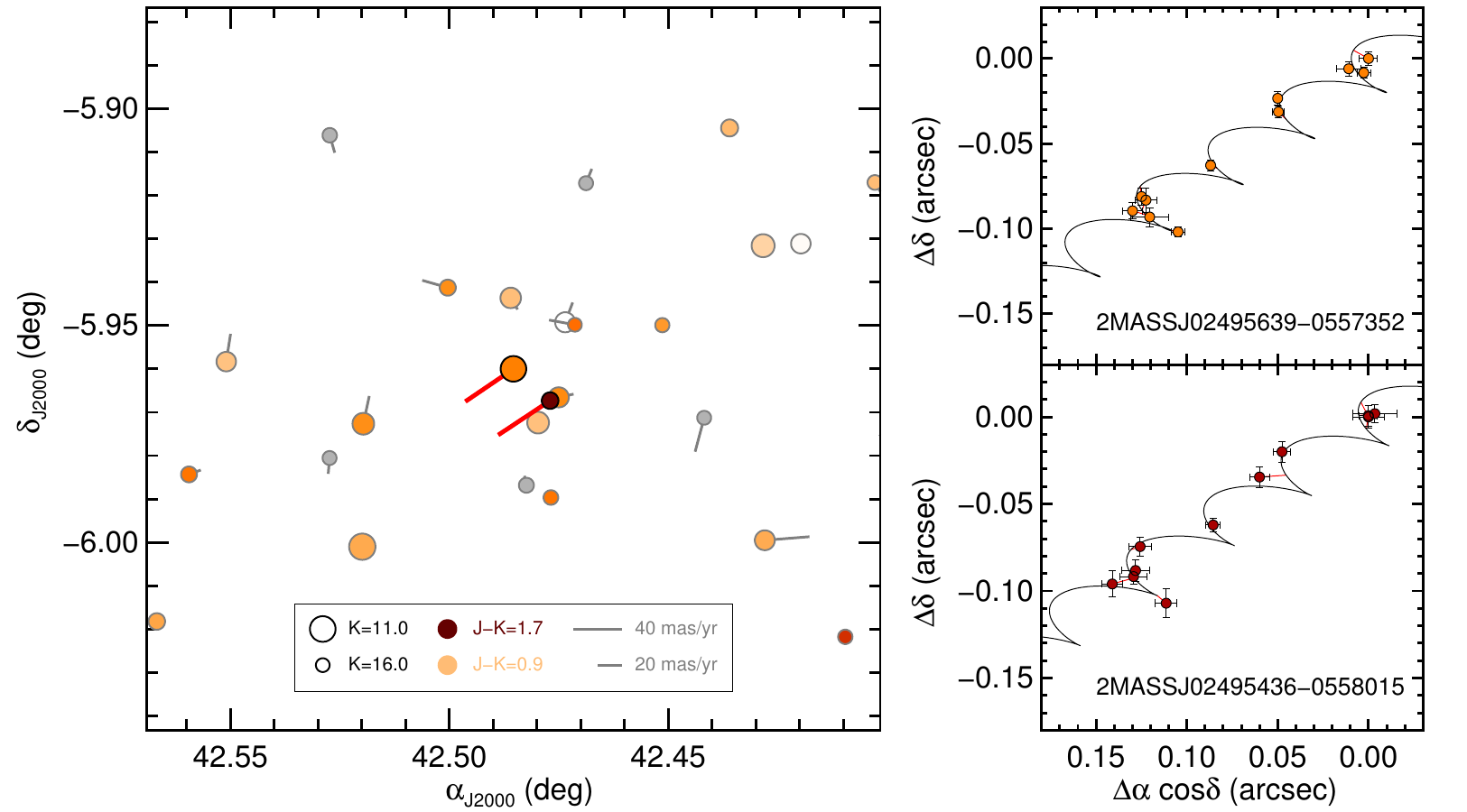}}
  \caption{\normalsize Left: All stars detected in our CFHT/WIRCam
    \Kn-band imaging and used in our astrometric analysis. Larger
    symbols indicate brighter stars, darker and redder symbols
    indicate redder $J-K$ colors based on 2MASS photometry (sources
    not in 2MASS are colored gray), and lines emanating from symbols
    indicate proper motion vectors where the tip of the line is the
    position 10$^3$\,yr from now. (Stars without lines have measured
    proper motions smaller than the symbol size.) The two objects in
    the center of the field with thick red proper motion vectors are
    the \bpic\ member \obj\ and our newly identified companion.
    Right: Relative astrometry of \obj\ (top) and the companion
    (bottom), where the origin corresponds to the earliest epoch. The
    best-fit proper motion and parallax solutions, computed separately
    for each object, are shown as black lines. The two objects have
    consistent proper motions and parallaxes (Table~\ref{tbl:plx})
    indicating that they are physically bound.  \label{fig:plx}}

\end{figure}

Unexpectedly, we found that one of the stars in the field had a proper
motion and parallax very similar to our intended target \obj\
(Figure~\ref{fig:plx}).  This other source has the Two-Micron All-Sky
Survey (2MASS) designation \objlongb, and its 2MASS photometry
indicates a very red color ($J-\Ks = 1.66\pm0.17$\,mag) that would be
consistent with being a later-type companion.  Table~\ref{tbl:cfht}
presents our measurements for both objects.  The median relative
astrometric error per epoch is 4.2\,mas for \obj\ (median SNR = 240)
and 6.0\,mas for the much fainter companion (median SNR = 15),
indicating that the uncertainty in the reference grid is setting the
astrometric noise floor, not the centroiding errors that scale as
$\propto {\rm FWHM}$/SNR.
The parallax and proper motion solutions for \obj\ and the companion
are given in Table~\ref{tbl:plx}. The relative proper motions of the
two objects are consistent within the errors (1.4$\sigma$), as are the
relative parallaxes (0.1$\sigma$).
We show in Section~\ref{sec:spt} that the companion is a young
L~dwarf, making it very improbable that it is an unrelated object in
the volume probed by our CFHT/WIRCam field of view (1.0\,pc$^3$ within
a distance limit of $<70$\,pc).  According to the 25-pc sample from
Best et al.\ (2018, in prep.), the space density of L0 and later
dwarfs of all ages is $\approx1\times10^{-2}$\,pc$^{-3}$, while for
young objects in the same spectral type range it is
$\approx6\times10^{-4}$\,pc$^{-3}$. Thus, the probability of the
companion being a chance alignment is $\ll1$\% even before considering
that it has consistent parallax and proper motion.  Therefore, we find
the two sources are physically associated.

We used the flux measurements reported by SExtractor ({\tt
  FLUX\_AUTO}) to compute relative photometry between \obj\ and the
companion at each epoch.  In order to examine photometric variations
in each object separately, we first computed the flux of each
component relative to a well-detected nearby reference star
(2MASS~J02495396$-$0557594, $\Ks = 13.28\pm0.04$\,mag).  In our
\Kn-band data, this reference star is $1.600\pm0.027$\,mag brighter
than the companion and $2.179\pm0.027$\,mag fainter than \obj\ itself.
These quoted flux ratio errors are the rms across all epochs, so
neither source appears to be more variable relative to the reference
star than the other. Computing the magnitude difference relative to
each other instead of the reference star gives
$\Delta\Kn = 3.780\pm0.032$\,mag, with $\chi^2=14.6$ (10 dof) using
the standard error at each epoch as quoted in Table~\ref{tbl:cfht},
which again is consistent with no variability above 0.03\,mag in \Kn\
band for either object.

\subsection{Keck/NIRC2 LGS AO  \label{sec:keck}}

\begin{figure} 
  \vskip -0.3in
  \centerline{
    \includegraphics[width=1.5in,angle=0]{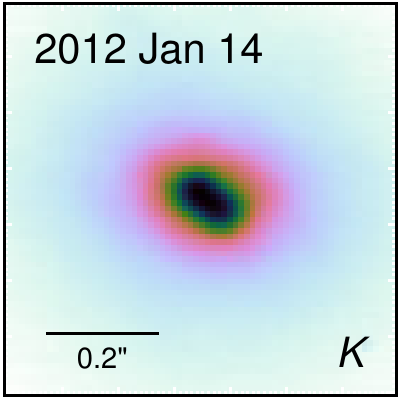} \hskip 0.1in
    \includegraphics[width=1.5in,angle=0]{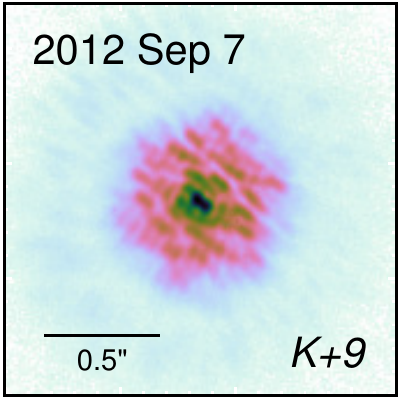} \hskip -0.1in 
    \includegraphics[width=1.5in,angle=0]{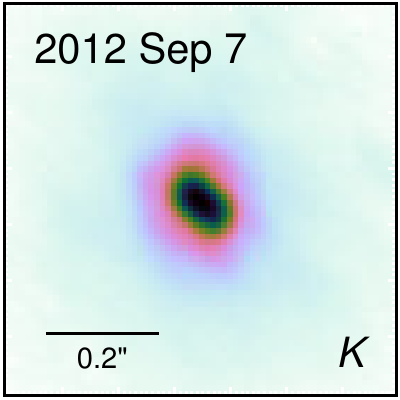} \hskip 0.1in
    \includegraphics[width=1.5in,angle=0]{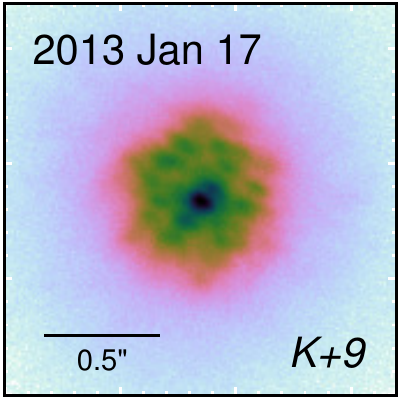}}
  \caption{\normalsize Keck/NIRC2 LGS AO images (left two) and 9-hole
    pupil-mask interferograms (right two) of \obj{AB}. All of these
    cutouts have been rotated for display purposes such that north is
    up and east is left and are shown with a square-root stretch. The
    interferogram cutouts show a larger area of the detector (i.e.,
    they are zoomed out) compared to the direct images. We are unable
    to derive astrometry from the imaging data because the binary is
    not cleanly resolved, but analysis of both masking observations
    results in significant binary detections and precise
    astrometry. For instance, in the 2012~Sep~7~UT data the imaging PSF
    is elongated at the same PA as the double peak in the center of
    the interferogram's PSF, and the masking analysis detects a binary
    with a separation of $44.4\pm0.2$\,mas and PA of
    $233\fdg3\pm0\fdg3$. \label{fig:keck}}
\end{figure}

We first observed the M6~dwarf \obj\ on 2012~Jan~28~UT using the laser
guide star adaptive optics (LGS AO) system at the Keck~II telescope
\citep{2004SPIE.5490..321B, 2006PASP..118..297W, 2006PASP..118..310V}.
We obtained several dithered $K$-band images with the facility
near-infrared camera NIRC2 and noted an elongation in the point-spread
function (PSF), but it was not clear if this was due to unstable AO
correction or a marginally resolved binary. On 2012~Sep~7~UT we
obtained data using the 9-hole nonredundant aperture mask installed
in the filter wheel of NIRC2 \citep{2006SPIE.6272E.103T}, in addition
to more imaging in which the PSF was elongated in a similar fashion as
the previous epoch.  We analyzed our masking data using the same
pipeline as in our previous papers
\citep[e.g.,][]{2008ApJ...678L..59I, 2009ApJ...699..168D,
  2015ApJ...805...56D, 2017ApJS..231...15D}.  The analysis indicated a
significant detection of a nearly equal-flux binary with the same PA
as the PSF elongation. In order to confirm the physical association of
this binary we obtained more masking data on 2013~Jan~17~UT and
recovered a detection at a similar separation, PA, and flux ratio.
Figure~\ref{fig:keck} shows examples of all of our imaging and masking
data.  In computing astrometry from our NIRC2 data, we adopt the
calibration from \citet{2010ApJ...725..331Y}, as appropriate for our
data taken during 2012--2013, which has a pixel scale of
$9.952\pm0.002$\,mas\,\perpix\ and an orientation for the detector's
$+y$-axis of $-0\fdg252\pm0\fdg009$ east of north.\footnote{In our
  past work \citep[e.g.,][]{2016ApJ...817...80D, 2017ApJS..231...15D}
  we reported PA values with a positive offset added to the header
  orientation, as prescribed by \citet{2010ApJ...725..331Y}. The
  offsets we used in the past were +0\fdg252 for
  \citet{2010ApJ...725..331Y} and +0\fdg262 for
  \citet{2016PASP..128i5004S}.  However, as discussed by
  \citet{2018AJ....155..159B} the sign of these offsets should be
  negative, not positive as stated in \citet{2010ApJ...725..331Y}.}

At discovery the separation of the binary was $44.4\pm0.2$\,mas, and
after 0.36\,yr it had moved inward to $40.1\pm0.2$\,mas. The total
motion of the secondary relative to the primary between the two epochs
was
$(\Delta\alpha\cos\delta, \Delta\delta) =
(+1.8\pm0.3,+5.0\pm0.4)$\,mas.  According to our CFHT parallax
solution for \obj, if the object in our Keck data were an unbound
background object with zero proper motion and parallax it would have
moved
$(\Delta\alpha\cos\delta, \Delta\delta) =
(+19.6\pm3.5,+19.5\pm1.1)$\,mas with respect to the primary.
Therefore, we conclude that the observed motion is consistent with
orbital motion as a physically bound binary system, since a background
object would require a finely tuned and high-amplitude proper motion
($\approx$20\,\masyr) to match our Keck LGS AO astrometry.
Table~\ref{tbl:keck} summarizes our measured astrometry and flux
ratios for this new binary \obj{AB}. We note that $\Delta{K}$ is
consistent within $\approx$0.01\,mag and within the quoted
uncertainties between the two epochs.

\subsection{IRTF/SpeX \label{sec:irtf}}

We obtained low-resolution near-IR (0.8--2.5\,\micron) spectra of
\obj~c on 2018~Feb~17~UT from the NASA Infrared Telescope Facility
(IRTF) located on Maunakea, Hawaii. Conditions were lightly cloudy
with 0\farcs9 seeing.  We used the facility near-IR spectrograph SpeX
\citep{1998SPIE.3354..468R} in prism mode with the 0\farcs8 slit. The
wavelength-dependent resolution with this slit ranges from
$R\approx50$ in $J$ band to $R\approx120$ in $K$ band.
We oriented the field to prevent other stars from landing on the slit.
This fixed PA did not correspond to the parallactic angle, but as we
discuss in Section~\ref{sec:spt} synthetic colors derived from our
spectrum agree well with 2MASS photometry, indicating that
wavelength-dependent slit losses were negligible.  We nodded the
object along the slit in an ABBA pattern with individual exposure
times of 180\,s, observed over an average airmass of 1.30.  We
observed the A0V star HD~18571 contemporaneously for telluric
calibration.
The total on-source integration time was 60~minutes. All spectra were
reduced using version 4.1 of the SpeXtool software
\citep{2003PASP..115..389V, 2004PASP..116..362C}.

\subsection{APO/TripleSpec \label{sec:apo}}

On 2018~Feb~27~UT, we obtained a moderate-resolution
($R \approx 3500$) near-IR spectrum of \obj~c using TripleSpec on
Apache Point Observatory's ARC~3.5~m telescope. TripleSpec
\citep{2004SPIE.5492.1295W} is a cross-dispersed spectrograph that
provides simultaneous wavelength coverage from 1.0--2.4\,\micron.
Conditions during our observations were clear with $\approx$1\farcs4
seeing, which was well matched to TripleSpec's 1\farcs1 slit. We
observed \obj~c for a total on-source integration time of 80~minutes
at an average airmass of 1.83. Over the course of our observations,
the orientation of the slit was continuously updated to the
parallactic angle to minimize atmospheric dispersion.  Immediately
following our observation, we observed the A0V star HD~25792 at an
airmass of 1.86 to correct for telluric absorption. All spectra were
reduced using a modified version of SpeXtool 4.1
\citep{2003PASP..115..389V, 2004PASP..116..362C}.

\section{Results}

\subsection{Spectral Classification \& Photometry \label{sec:spt}} 

\begin{figure} 
  \centerline{\includegraphics[width=6.5in,angle=0]{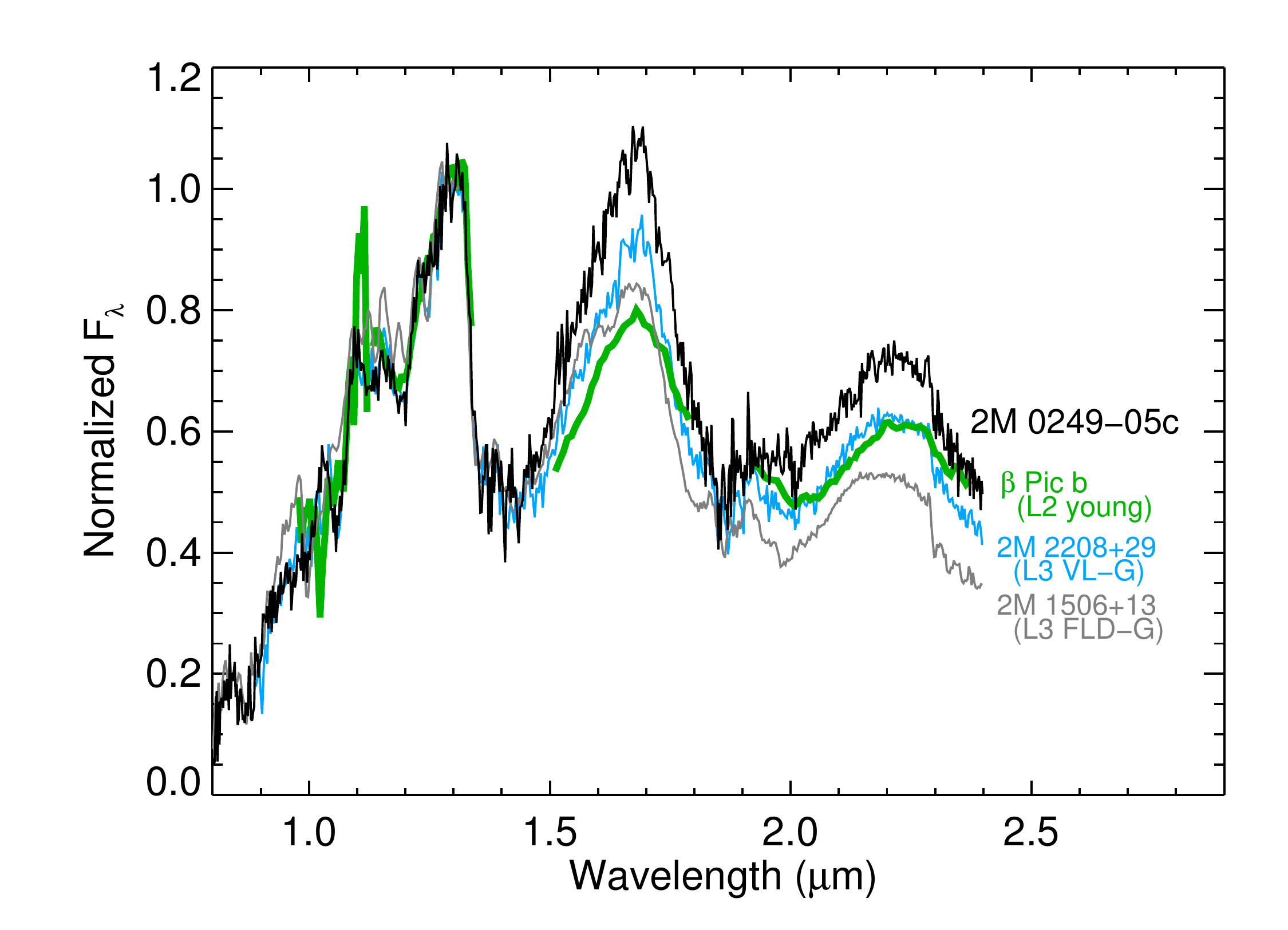}}
  \caption{\normalsize Near-IR IRTF/SpeX spectrum of \obj~c compared
    to the near-IR field standard 2MASS~J1506+1321
    \citep{2010ApJS..190..100K}, the young (\vlg) standard
    2MASS~J2208+2921 \citep{2013ApJ...772...79A}, and the young
    exoplanet \bpic~b \citep{2017AJ....153..182C}.  The spectra are
    normalized at the peak region in the $J$-band
    (1.26--1.31~\micron).  For the two standards, the SpeX data were
    taken with the 0\farcs5 slit (wavelength-dependent
    $R\approx80$--200), and our SpeX spectrum of \obj~c was taken with
    the 0\farcs8 slit ($R\approx50$--120).  The \bpic~b data come from
    the Gemini Planet Imager and have a spectral resolution ranging
    from $R\approx35$ (at $Y$~band) up to $R\approx75$ (at $K$~band),
    too coarse for AL13 gravity classification.  For \bpic~b, two
    strongly discrepant data points around 1.30~\micron\ have been
    removed for plotting purposes.  \label{fig:spectra}}

\end{figure}

\begin{figure} 
  \hskip -0.2in
  \includegraphics[width=10in,angle=0]{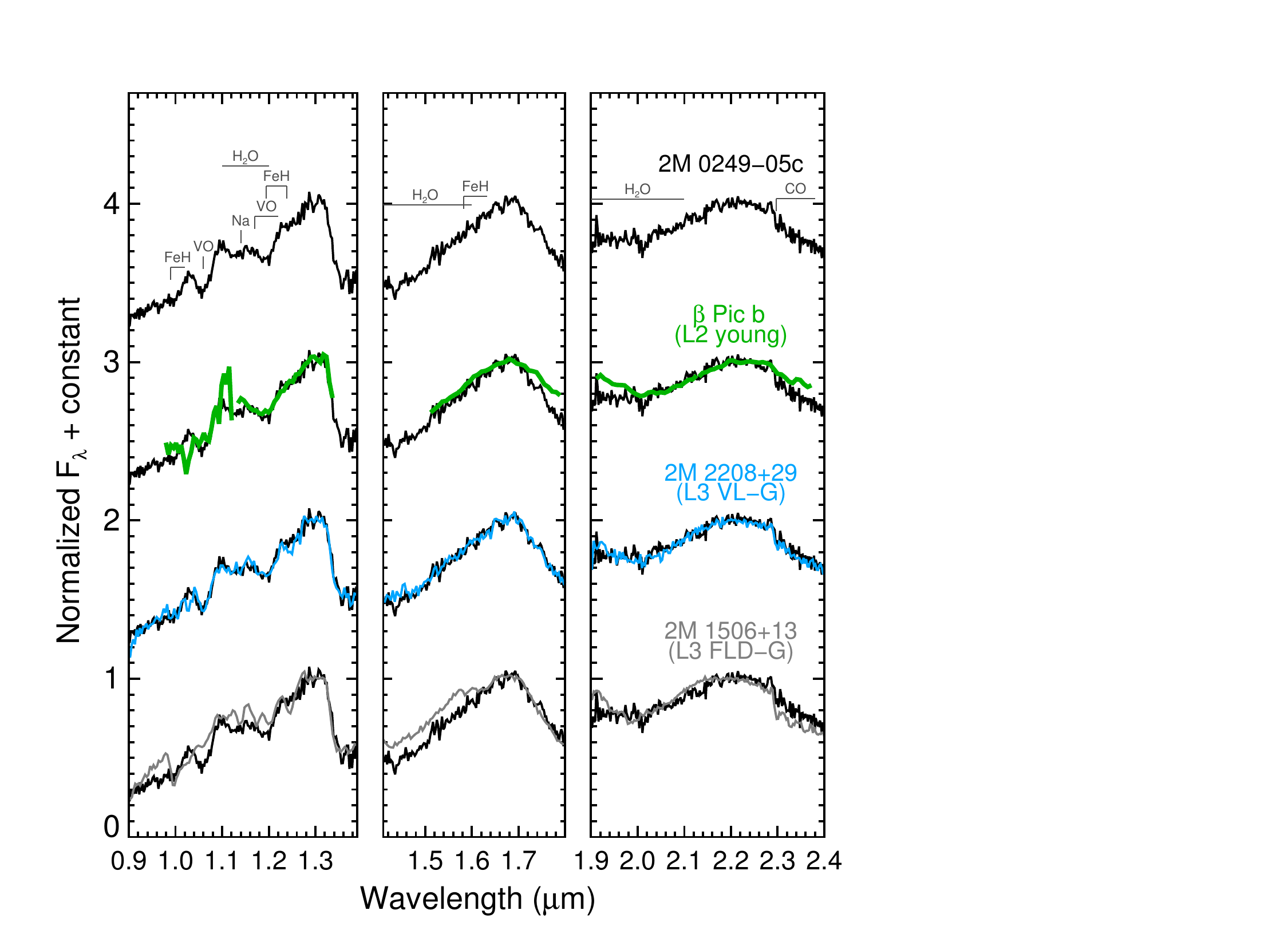}
  \caption{\normalsize Comparison of the same spectra as in
    Figure~\ref{fig:spectra}, now with each bandpass separately
    normalized.  The \obj~c spectrum is plotted four times.  The
    L3~\vlg\ standard 2MASS~J2208+2921 provides an excellent match in
    all bands, with \bpic{b} also being quite
    similar. \label{fig:3bands}}
\end{figure}

\begin{figure} 
  \hskip -0.2in
  \centerline{\includegraphics[width=6.5in,angle=0]{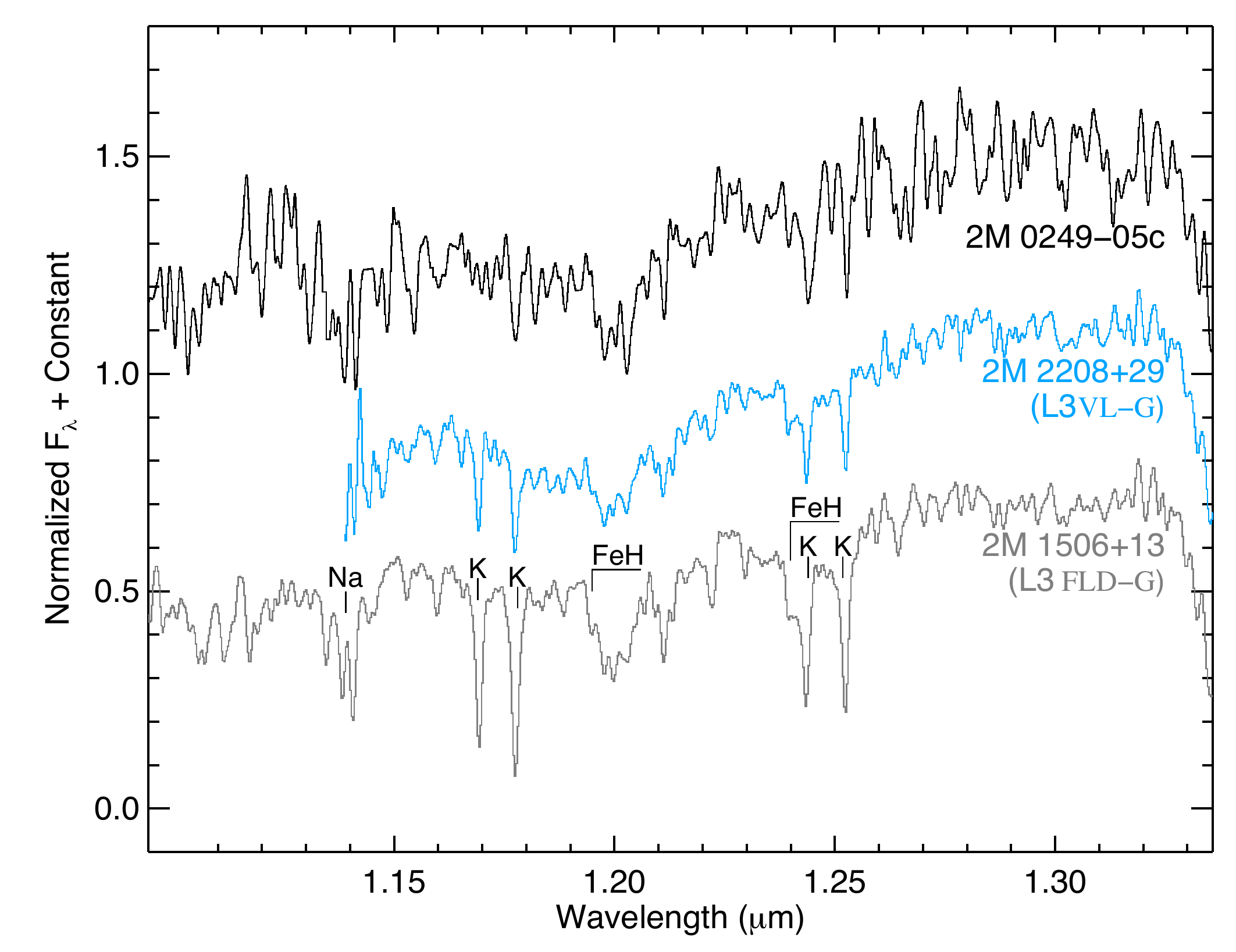}}
  \caption{\normalsize $J$-band APO/TripleSpec spectrum of \obj~c
    compared to the spectra of the near-IR field standard
    2MASS~J1506+1321 \citep{2005ApJ...623.1115C} and the young (\vlg)
    object 2MASS~J2208+2921 \citep{2017ApJ...838...73M}. All spectra
    have been smoothed to $R\approx1200$, are normalized by their
    median flux from 1.27--1.31\,\micron, and then offset by a
    constant. The spectrum of \obj~c shows the weak \ion{K}{1},
    \ion{Na}{1}, and FeH absorption features indicative of a young,
    low-gravity object. \label{fig:apo}}
\end{figure}

Figures~\ref{fig:spectra},~\ref{fig:3bands},~and~\ref{fig:apo} show
the spectrum of \obj~c compared to objects of similar near-IR spectral
type. The low-gravity nature of the object is seen clearly in the
$H$-band continuum shape (triangular compared to field objects),
$K$-band continuum shape (redder continuum peak and different
curvature of the blueward continuum), VO 1.08~\micron\ absorption
(stronger), and FeH 0.99~\micron\ absorption (weaker) as discussed,
e.g., in \citet[][hereinafter AL13]{2013ApJ...772...79A}. To assign a
spectral type and to assess the gravity for \obj~c, we follow the
near-IR classification methods of AL13. This approach uses a
combination of qualitative visual typing with quantitative measurement
of flux indices to determine a spectral type.  The AL13 approach then
determines a gravity classification using flux indices and equivalent
widths of gravity-sensitive features.

For visual typing, we compare our SpeX $J$- and $K$-band spectra to
near-IR spectroscopic standards for field (high-gravity [\fldg])
objects from \citet{2010ApJS..190..100K} and for young (low-gravity
[\vlg]) objects from AL13. Following the prescription of AL13, in each
bandpass we normalize the fluxes of \obj~c and the spectroscopic
standards prior to visual comparison.\footnote{The AL13 approach of
  normalizing each bandpass separately prior to visual comparison,
  rather than normalizing the entire near-IR spectrum, is conceptually
  identical to the classification system recently proposed by
  \citet{2018AJ....155...34C}. Their study does include $H$~band for
  visual classification, which AL13 does not, and also has a few
  differences in the spectroscopic standards.} The near-IR spectrum of
\obj~c matches the L3~\vlg\ standard 2MASSW~J2208136+292121
(hereinafter \objtwo) very well, even for the $H$-band spectrum though
this was not used for typing.

For index-based analysis, we use the approach of
\citet{2016ApJ...821..120A}, which calculates the AL13 indices and
includes a Monte Carlo estimation of the measurement errors.
Combining spectral types calculated from AL13's four gravity-sensitive
indices (L$2.4\pm1.2$, L$2.1\pm1.0$, L$0.5\pm1.2$, and L$2.0\pm1.0$)
with our visual classification of the SpeX $J$- and $K$-band spectra
(L$3\pm1$ and L$3\pm1$, respectively), we assign a spectral type of
L$2\pm1$.

From the low resolution SpeX spectrum, we find an AL13 gravity score
of 1222, which represents the gravity inferred from four spectral
features, leading to a gravity classification of \vlg.  We also used
our moderate-resolution TripleSpec spectrum to calculate the AL13
low-resolution gravity-sensitive indices as well as the additional
AL13 indices and equivalent widths available at moderate resolution.
We find an AL13 gravity score of 2222 for our TripleSpec spectrum,
confirming the \vlg\ classification determined from our lower
resolution SpeX spectrum.  We assign a final classification of
L$2\pm1$~\vlg.

We also use our spectrum of \obj~c along with the published
integrated-light SpeX spectrum of \obj{AB} from
\citet{2017AJ....154...69S} to synthesize photometry on the MKO
system.  We first synthesize offsets for both objects between the MKO
and 2MASS photometric systems in each bandpass, as well as the offset
between broadband $K$ and the narrow \Kn\ bandpass used in our
CFHT/WIRCam imaging.  We used 2MASS photometry
\citep{2003tmc..book.....C} to flux calibrate the spectrum of
\obj{AB}, and then we used our CFHT flux ratio
($\Delta\Kn = 3.780\pm0.032$\,mag) to flux calibrate the spectrum of
\obj~c.  In this process, we checked our synthesized 2MASS $JH\Ks$
magnitudes against the 2MASS photometry of \obj~c and found good
agreement, $p(\chi^2) = 0.35$, but with our synthesized photometry
having much smaller errors.  The resulting synthesized $JHK$
photometry on the MKO system for both objects is given in
Table~\ref{tbl:prop}.  As in our previous work with synthesized
photometry \citep{2012ApJS..201...19D}, we consider the errors on
photometric system offsets negligible compared to the uncertainties in
2MASS photometry, and we adopt 0.05\,mag errors on synthesized
magnitudes when no direct photometry is available.

\subsection{Bolometric Fluxes \label{sec:fbol}} 

In order to ultimately derive physical properties for the components
of the \obj\ system we must first estimate their bolometric fluxes. We
use the procedure from \citet{2015ApJ...804...64M}, which we briefly
summarize here. For both \obj{AB} (in integrated light) and \obj~c, we
compiled optical and IR photometry from SDSS-DR14
\citep{2018ApJS..235...42A}, 2MASS, and the \textsl{Wide-field
  Infrared Survey Explorer} \citep[\WISE;][]{2010AJ....140.1868W}. We
also used the MKO $K$-band photometry of \obj~c from
Section~\ref{sec:spt} that is based on our CFHT/WIRCam imaging. For
each object we compared all available photometry to synthetic
magnitudes computed either from observed, template, or model spectra.
For \obj{AB} we used our IRTF/SpeX spectrum and a template optical
spectrum of the M6~\vlg\ object 2MASS~J03363144$-$2619578 obtained
with SNIFS (Mann et al., in preparation).  For \obj~c we used the
combination of our IRTF/SpeX spectrum, an optical spectrum from SDSS,
and a BT-Settl model \citep{2011ASPC..448...91A} in regions not
covered by the empirical spectra.
To compute synthetic magnitudes from each spectrum we used appropriate
filter profiles and zero-points \citep[e.g.,][]{2003AJ....126.1090C,
  2011ApJ...735..112J}. Spectra were then scaled to match all
available photometry, using the overlapping wavelengths of the IR and
optical spectra (0.75--0.85\,\micron) as an additional constraint.

\begin{figure} 
  \vskip -0.3in
  \centerline{\includegraphics[width=4.5in,angle=0]{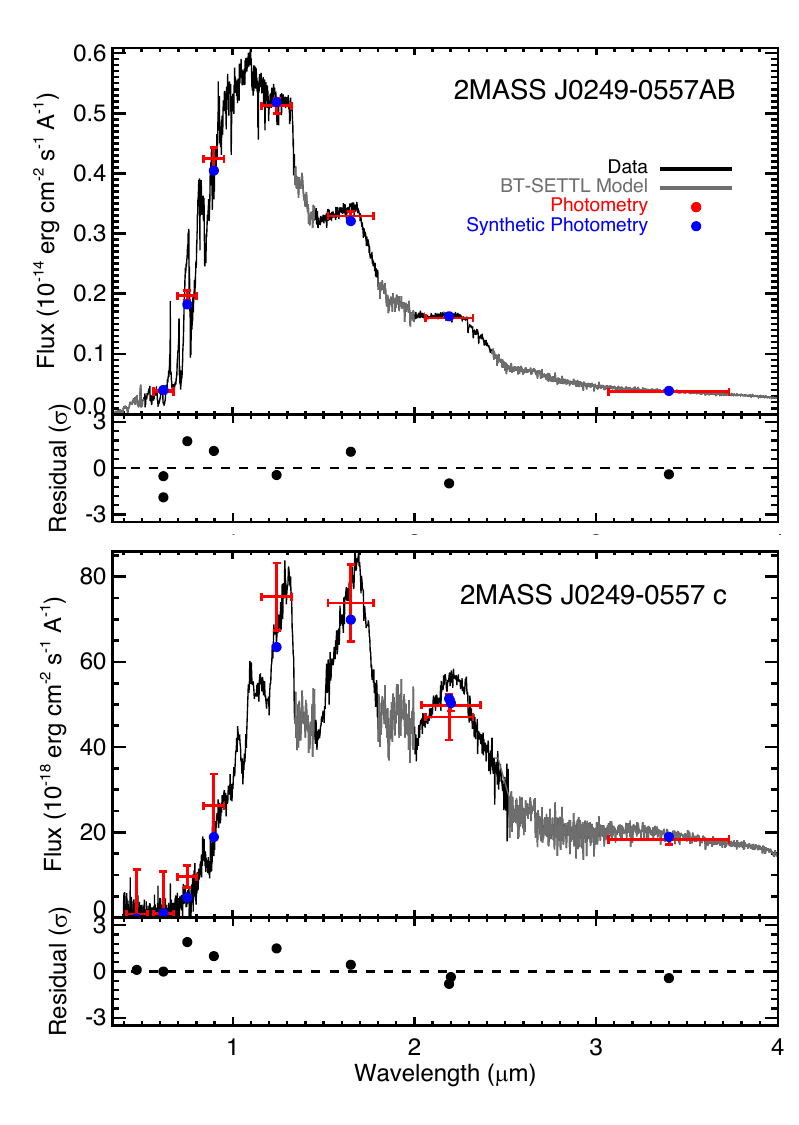}}
  \caption{\normalsize Absolutely flux-calibrated spectra that we use
    to compute bolometric fluxes (\fbol).  Black indicates directly
    observed spectra, and gray indicates wavelength regions that are
    likely to have high telluric contamination or that are beyond the
    observations, which we have filled in using BT-Settl atmospheric
    models. Red points are literature photometry, where $y$-axis error
    bars correspond to reported measurement uncertainties and $x$-axis
    error bars indicate the width of the filter. Blue points show
    synthetic photometry computed from the displayed spectrum.  The
    bottom panel shows residuals (observed minus synthetic photometry)
    in units of standard deviations. \label{fig:fbol}}
\end{figure}

Figure~\ref{fig:fbol} shows final calibrated spectra. To compute
bolometric fluxes (\fbol) we integrated over these joined and
absolutely calibrated spectra. We derived \fbol\ errors accounting for
uncertainties in the spectral flux calibration, filter zero-points,
and Poisson errors in the observed photometry and spectra, yielding
final values of
$(1.35\pm0.07) \times 10^{-12}$\,erg\,cm$^{-2}$\,s$^{-1}$ for \obj~c
and $(6.56\pm0.29) \times 10^{-11}$\,erg\,cm$^{-2}$\,s$^{-1}$ for
\obj{AB} in integrated light.
Table~\ref{tbl:prop} summarizes these results in terms of apparent
bolometric magnitudes (\mbol) so that future improvements in distance
measurements can be readily applied. Our integrated-light magnitude
for \obj{AB} of $\mbol=13.96\pm0.05$\,mag is in good agreement with
the value of $13.92\pm0.02$\,mas determined from photometry alone by
\citet{2017AJ....154...69S}.

\subsection{Membership Assessment for \obj \label{sec:memb}}

Our new parallax and independently measured proper motion allow us to
reexamine the membership of \obj{AB} in the \bpic\ moving group, as
the original analysis by \citet{2017AJ....154...69S} used a less
precise proper motion, did not have a parallax, and did not know it
was an unresolved binary.
In fact, even our proper motion measurement could be influenced by
photocenter motion due to the binary orbit of \obj{AB}.  Over short
time baselines, long-term orbital motion can cause systematic offsets
in measured proper motions (e.g., see Section~2.4.1 of
\citealp{2012ApJS..201...19D}), while parallaxes are not commonly
affected systematically.  Fortunately, the companion has similar
proper motion precision as \obj{AB} but is less likely harbor unknown
systematic errors due to orbital motion, as it is not known to be a
binary and is marginally fainter than average for its spectral type
(Section~\ref{sec:cmd}).  Therefore, in the following kinematic
analysis we use the proper motion of \obj~c but the more precise
parallax of \obj{AB}.

\subsubsection{Binary Influence on the RV \label{sec:rv-bin}}

We consider the possibility that the unresolved binarity of \obj{AB}
may have influenced the \citeauthor{2017AJ....154...69S} radial
velocity (RV) measured in optical integrated light.  The velocity
difference of the two components was not large enough relative to
\vsini\ to appear as a double-lined spectroscopic binary, but the line
centroids could have been shifted by the binary orbit.  If this were
an exactly equal-flux, equal-mass binary then the spectral lines would
broaden slightly while remaining centered at the system velocity.  But
for an arbitrary flux ratio and mass ratio the flux-weighted centroid
shift of the spectral lines away from the system velocity is
$$\Delta{\rm RV}_{\rm orb}\left(\frac{F_{\rm A}}{F_{\rm tot}}\frac{M_{\rm B}}{M_{\rm tot}}-\frac{F_{\rm B}}{F_{\rm tot}}\frac{M_{\rm A}}{M_{\rm tot}}\right),$$
where $\Delta{\rm RV}_{\rm orb}$ is the RV difference between the two
components at a given epoch.  As described in Section~\ref{sec:mass},
we estimate a mass ratio of $M_{\rm B}/M_{\rm A} = 0.9$ from
evolutionary models at the age of the \bpic\ moving group. (The mass
ratio would be negligibly different at older field ages.)  Using our
Keck infrared flux ratio of $\Delta{K} = 0.123\pm0.005$\,mag with the
BT-Settl evolutionary model magnitudes \citep{2011ASPC..448...91A}, we
estimate an $r$-band flux ratio of $F_{\rm B}/F_{\rm A} = 0.75$
(0.31\,mag), and this is also essentially the same for young and old
ages.  Thus, the factor by which $\Delta{\rm RV}_{\rm orb}$ must be
multiplied to compute the expected shift in the integrated-light RV
(i.e., the term in parentheses in the equation above) is 0.045
assuming \bpic\ membership.\footnote{We note that our estimate of RV
  systematic errors neglects the fact that slit losses can cause a
  binary to experience RV shifts depending on how the slit is centered
  with respect to the individual components.  However, this
  approximation is justified here because the 0\farcs5 slit used by
  \citeauthor{2017AJ....154...69S} and typical seeing values are
  $\gtrsim$10$\times$ larger than the binary separation.}

We consider the possibilities of low and high eccentricity orbits and
conservatively assume an edge-on orbit, which would produce maximal
RVs.  As described below, the detection of lithium implies that the
system must be younger than $\approx$100\,Myr, corresponding to masses
of 0.08--0.11\,\Msun\ for the components of \obj{AB}, depending on the
age, so we assume a system mass of 0.2\,\Msun\ to convert semimajor
axis to orbital period.  To estimate the semimajor axis from the
observed projected separation, we use the conversion factors
calculated by \citet{2011ApJ...733..122D} for very low-mass binaries.
Because this binary was discovered near the resolution limit of our
Keck imaging, we use the value of $a/\rho = 0.85$ corresponding to
severe discovery bias.  Thus our measured separation of
$2.17\pm0.22$\,AU implies a semimajor axis of $1.8\pm0.3$\,AU and
orbital period of 5.4\,yr for a system mass of 0.2\,\Msun.  In this
case the median $\Delta{\rm RV}_{\rm orb}$ is 6.6\,\kms\ for low
eccentricity ($e=0.2$) and 2.9\,\kms\ for high eccentricity ($e=0.8$).
The maximum possible $\Delta{\rm RV}_{\rm orb}$ for these orbits are
12\,\kms\ and 30\,\kms, respectively. A fractional shift of
0.045$\times\Delta{\rm RV}_{\rm orb}$ implies typical deviations from
systemic velocity in the integrated light spectral lines of
0.30\,\kms\ (up to 0.55\,\kms) for low eccentricity and 0.13\,\kms\
(up to 1.3\,\kms) for high eccentricity, depending on orbital phase.
The integrated-light RV of $14.42\pm0.44$\,\kms\ from
\citet{2017AJ....154...69S} was measured on 2010~Dec~31~UT, 1.69\,yr
prior to our first Keck LGS AO astrometry.  Thus we cannot rule out a
scenario in which \obj{AB} is an eccentric binary that would have been
going through periastron passage (i.e., maximum
$\Delta{\rm RV}_{\rm orb}$) at the RV measurement epoch.  We therefore
conservatively assume an uncertainty of 1.3\,\kms\ on the system
velocity as measured by the integrated-light RV.\footnote{We note that
  as a \bpic\ moving group member we would expect somewhat smaller
  integrated-light RV excursions for \obj{AB} than estimated above
  because a younger age corresponds to lower masses for the components
  of \obj{AB} and thereby a longer orbital period for a given
  semimajor axis.  As we derive in Section~\ref{sec:mass}, a smaller
  system mass of $\approx$0.1\,\Msun\ is predicted from models,
  implying a longer orbital period of 8\,yr and thereby smaller median
  RV deviations of 0.10\,\kms\ (up to 1.0\,\kms) for an eccentric
  orbit. However, a change in RV uncertainty from 1.3\,\kms\ to
  1.0\,kms\ has a negligible impact on our following analysis.}

\subsubsection{Reaffirming ACRONYM Membership \label{sec:acronym}}

Combining the RV from the \citet{2017AJ....154...69S} ACRONYM survey
with our absolute parallax and proper motion, we find
$(U,V,W) = (-10.7\pm0.9,-12.8\pm1.4,-9.8\pm1.1)$\,\kms\ and
$(X,Y,Z) = (-28\pm3,-0.68\pm0.07,-40\pm4)$\,pc.  
Despite our parallax distance ($48.9^{+4.4}_{-5.4}$\,pc) being
somewhat smaller than the kinematic distance of 60\,pc used in the
analysis of \citeauthor{2017AJ....154...69S}, our $XYZ$ values agree
within 0.7--1.0$\sigma$ of their values
$(X,Y,Z) = (-35\pm4,-0.80\pm0.08,-49\pm5)$\,pc.  Here we assume that
the $XYZ$ uncertainties of \citeauthor{2017AJ....154...69S} are
dominated by kinematic distance uncertainty, which we estimate to be
10\% based on the fractional error in the proper motion they used
$(44.6\pm4.1,-35.0\pm4.1)$\,\masyr.
Compared to their $(U,V,W) = (-10.6,-16.2,-10.0)$\,\kms, only the $V$
component is more than 0.2$\sigma$ different from our own
measurements.  Examining the covariance between our input measurements
and output velocities indicates that this discrepancy in $V$ is almost
entirely due to the difference in the declination component of our
CFHT proper motion $\mu_{\delta} = -32.0\pm2.1$\,\masyr\ for \obj~c
compared to $-35.0\pm4.1$\,\masyr\ for \obj{AB} in
\citet{2017AJ....154...69S}.  These two independent measurements are
consistent within 0.7$\sigma$; therefore, we conclude that our updated
$UVW$ velocity is consistent within the 1$\sigma$ uncertainties of the
kinematic data used by \citet{2017AJ....154...69S}. By extension, we
expect that their assessment of \obj\ as a likely member of the \bpic\
moving group would remain unchanged using our new proper motion, but
we now consider membership in more detail given our addition of a
parallax.

Because the \bpic\ moving group is spread over thousands of square
degrees, both the directly observable kinematics of members (proper
motion and RV) and contamination due to the field population will vary
widely over the sky. Achieving a highly complete group census requires
casting a wide net in kinematic space, but not so wide as to become
unacceptably contaminated by field objects. We consider these two
competing effects in the following.

First, to estimate the completeness of selecting \bpic\ group members
using various kinematic criteria, we created a Monte Carlo population
of simulated members at the sky position and distance of \obj.  For
the kinematics of the \bpic\ moving group, we consider two velocity
ellipsoids derived from slightly different membership lists.  One is
the Gaussian ellipsoid derived by \citet{2014MNRAS.445.2169M}, which
has a mean velocity of $(U,V,W) = (-10.9,-16.0,-9.2)$\,\kms\ and
intrinsic velocity dispersions along these axes of
$(1.5,1.4,1.8)$\,\kms, respectively. This is based on the classic
membership list of 26~stars from \citet{2004ARA&A..42..685Z} plus four
additional high-probability members from
\citet{2013ApJ...762...88M}. We also consider an ellipsoid based on a
somewhat larger membership list of 57~stars from
\citet{2018MNRAS.475.2955L} that was derived from a uniform assessment
of all potential \bpic\ candidates in the literature at the time,
selecting only the highest probability members. The mean velocity of
the \citet{2018MNRAS.475.2955L} ellipsoid,
$(U,V,W) = (-10.5,-15.9,-9.1)$\,\kms, is nearly identical to that of
\citet{2014MNRAS.445.2169M} but with dispersion axes that are rotated
to match the covariances in the data as fit by three Euler angles.
Our $UVW$ for \obj{AB} places it 3.5\,\kms\ away from the mean
velocity of the \bpic\ moving group using either the
\citet{2014MNRAS.445.2169M} or \citet{2018MNRAS.475.2955L} results.

\begin{figure} 
  \hskip -0.2in
  \centerline{
    \includegraphics[width=3.2in,angle=0]{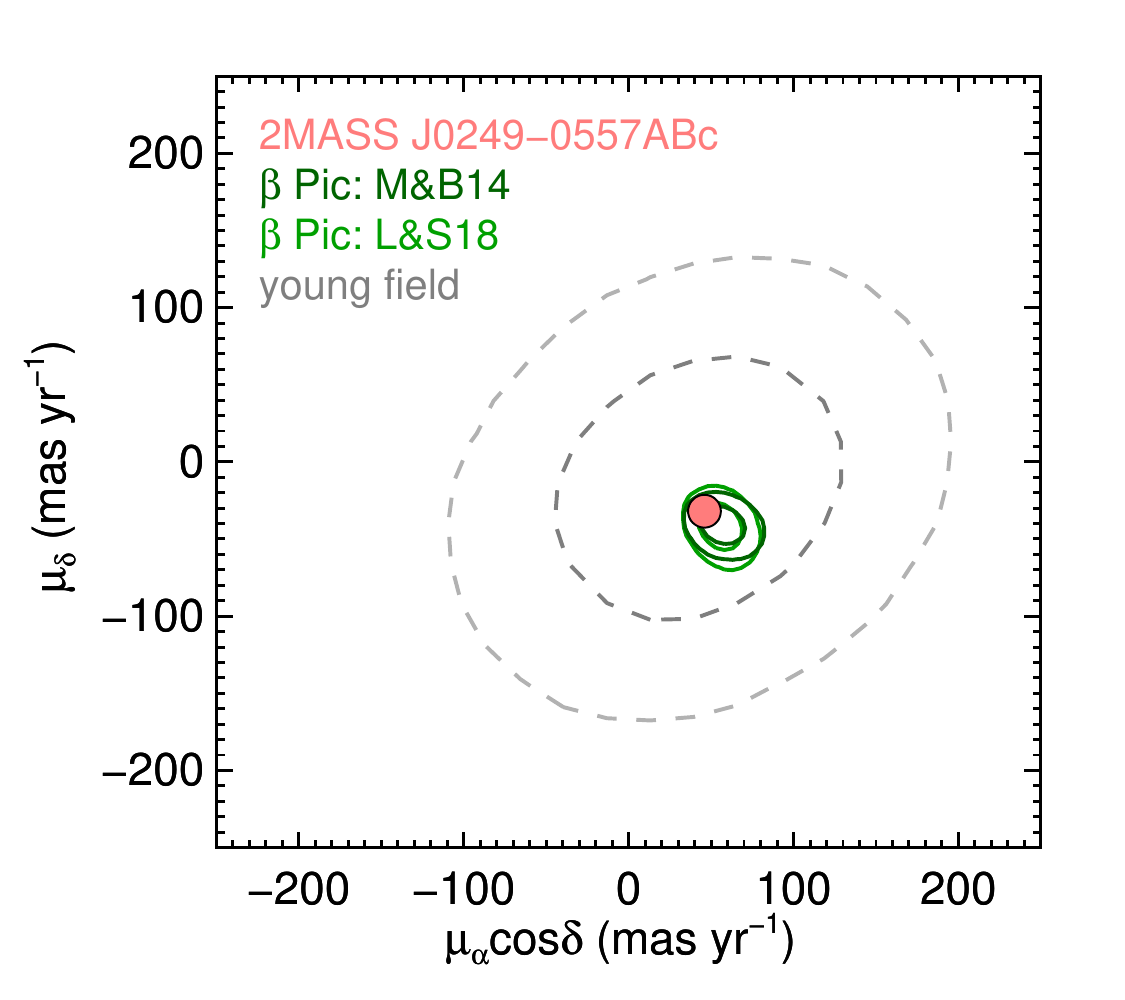}
    \includegraphics[width=3.2in,angle=0]{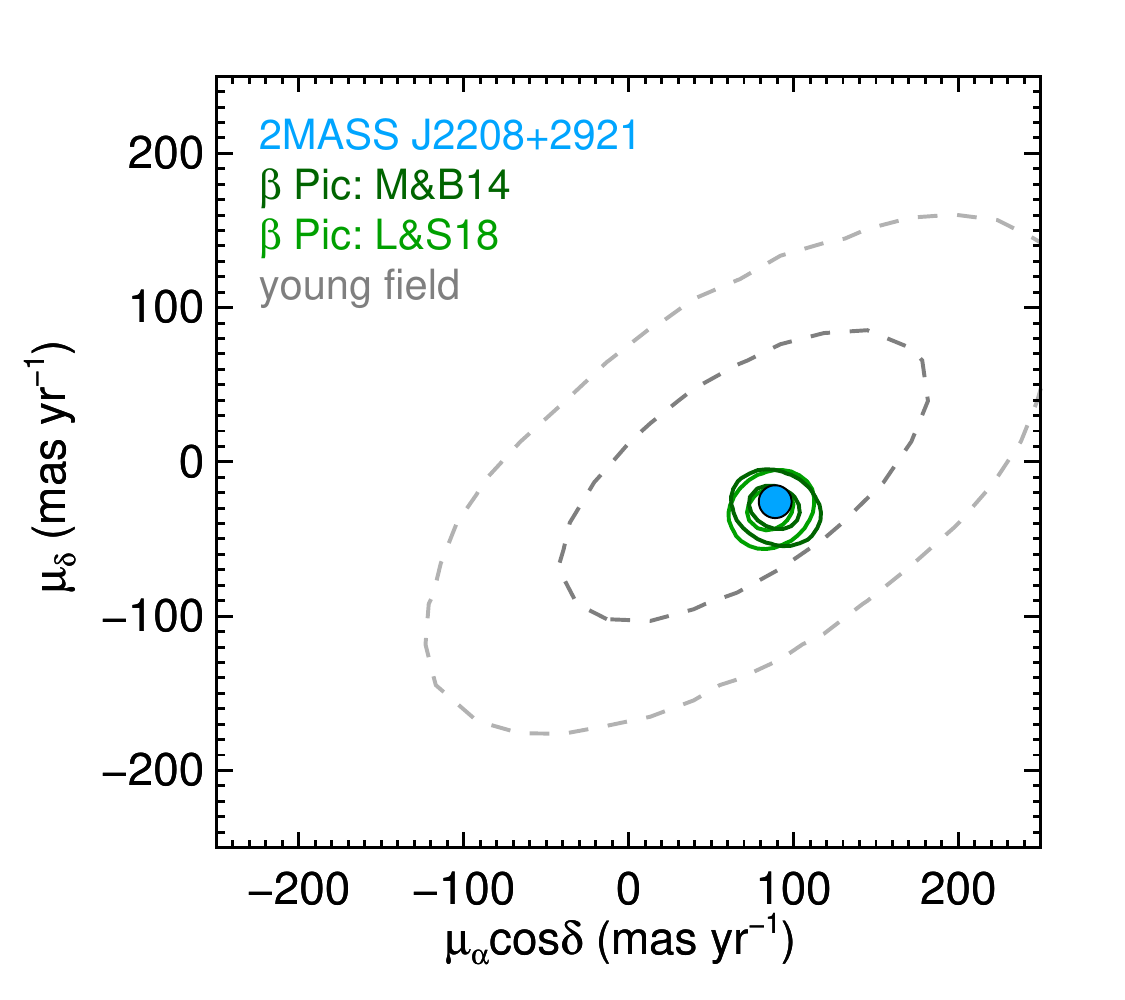}}
  \caption{\normalsize Measured proper motions for the \obj\ system
    (left) and \objtwo\ (right) shown alongside various populations
    that we simulated in $UVW$ space and then projected into proper
    motion space using the measured parallaxes.  Two different
    velocity ellipsoids are shown for the \bpic\ moving group from
    \citet{2014MNRAS.445.2169M} and \citet{2018MNRAS.475.2955L}, which
    give very similar results.  The young field population is from the
    Besan\c{c}on model of the Galaxy \citep{2003A&A...409..523R}.  For
    these populations of simulated objects we show 2-d contours
    containing 68.3\% and 95.4\% of the objects in proper motion
    space.  The measured proper motions of both systems are consistent
    with \bpic\ membership but also with the broadly distributed young
    field population. (This is for display purposes only as our
    analysis is based on the full 3D kinematics including
    RV.)  \label{fig:pm-bpmg}}
\end{figure}

To properly account for all covariances, we project the $UVW$
velocities of the simulated \bpic\ population into proper motions and
RVs using the sky coordinates, parallax, and corresponding measurement
uncertainties of the \obj\ system. In proper motion--RV space we can
more clearly investigate observational selection effects.
Figure~\ref{fig:pm-bpmg} shows the projection of the 3D velocity
ellipsoids into 2-d proper motion space. For display purposes,
Figure~\ref{fig:pm-bpmg} also shows contours corresponding to the
young field population ($<$150\,Myr) as simulated in the Besan\c{c}on
model of the Galaxy \citep{2003A&A...409..523R}.  As expected, the
field population spans a large amount of parameter space, encompassing
the entirety of the \bpic\ moving group and \obj.

For every simulated \bpic\ group member and \obj\ itself we compute
the 3D distance in $(\mu_{\alpha}\cos\delta,\mu_{\delta},{\rm RV})$
space from the mean velocity. In general, we quote 3D distances in
units of $\sigma$, i.e., normalized along each principal component
axis by the standard deviation in that direction (a.k.a., the
Mahalanobis distance). In 3D space, the usual Gaussian confidence
intervals do not correspond to integer units of $\sigma$, but rather
68.3\% of the distribution is contained within 1.88$\sigma$, 95.4\%
within 2.83$\sigma$, and so on. \obj\ is fairly close to the mean
velocity of the \bpic\ moving group, only 1.46$\sigma$ and
1.32$\sigma$ away from the ellipsoids of \citet{2014MNRAS.445.2169M}
and \citet{2018MNRAS.475.2955L}, respectively. Among simulated \bpic\
group members, most of them (54\% and 63\%, respectively) are more
distant than \obj\ from the ellipsoid means. Thus, \obj\ would pass
any reasonably inclusive kinematic criteria for membership in the
\bpic\ moving group using this approach.

To estimate the probability that \obj\ could be a field interloper
that happens to share the kinematics of the \bpic\ moving group, we
consider the ACRONYM search in which it was originally identified
\citep{2017AJ....154...69S}. The first step in this search was to
select candidates based on astrometry and photometry, using proper
motions to estimate a kinematic distance (the distance required to
minimize the difference between the measured and expected proper
motion of a \bpic\ member at the given RA and Dec) and SEDs to
estimate spectral types. Of the $4.5\times10^3$ objects with proper
motions consistent with \bpic\ kinematics, only 104 objects with
estimated spectral types of K7--M9 were consistent with being young on
the H-R diagram and thus selected for spectroscopic follow up. The
latest-type sources were first screened for signs of low gravity using
low-resolution IR spectra. High-resolution optical spectroscopy was
obtained for all remaining objects to measure RVs and look for
H$\alpha$ emission and \ion{Li}{1} absorption. This resulted in 91
objects with RV measurements, including both \bpic\ members and field
contaminants. For each object, \citet{2017AJ....154...69S} computed
the expected RV for \bpic\ motion, and the difference from the
measured value ($\Delta{\rm RV}_{\rm BPMG}$) was used, along with
other youth indicators, in assigning final memberships. Here we use
the $\Delta{\rm RV}_{\rm BPMG}$ distribution of the objects that
\citet{2017AJ....154...69S} classified as nonmembers to estimate the
fraction of field interlopers that would pass both the initial proper
motion and final RV selection.

We assume that both populations (members and nonmembers) found in the
ACRONYM search can be approximated as Gaussians in
$\Delta{\rm RV}_{\rm BPMG}$. The nonmembers should be distributed
widely in $\Delta{\rm RV}_{\rm BPMG}$ while members cluster tightly
around zero. \citet{2017AJ....154...69S} used a threshold in
$\Delta{\rm RV}_{\rm BPMG}$ of 5.4\,\kms\ to select members,
corresponding to a 3$\sigma$ cut given the velocity dispersion of
1.8\,\kms\ used in their analysis for the \bpic\ moving group. We
assume that the subset of 32 objects that did not pass their
$\Delta{\rm RV}_{\rm BPMG}$ cut or lacked H$\alpha$ emission (as
expected at the age of \bpic) represent the contaminant population,
and these objects have a mean and standard deviation in
$\Delta{\rm RV}_{\rm BPMG}$ of $13\pm46$\,\kms. In contrast, the 52
objects identified as members have a $\Delta{\rm RV}_{\rm BPMG}$ mean
and standard deviation of $0.3\pm2.4$\,\kms. (Here we have excluded
seven objects with ambiguous status, mostly spectroscopic binaries
where the RV likely has a systematic orbital offset.) To compute a
false-alarm rate, we combine these two Gaussians into a single
probability distribution, normalized according to $52/84=62$\% members
(centered at $\Delta{\rm RV}_{\rm BPMG} = 0$) and $32/84=38$\%
contaminants.

\begin{figure} 
  \hskip -0.2in
  \centerline{\includegraphics[width=6.0in,angle=0]{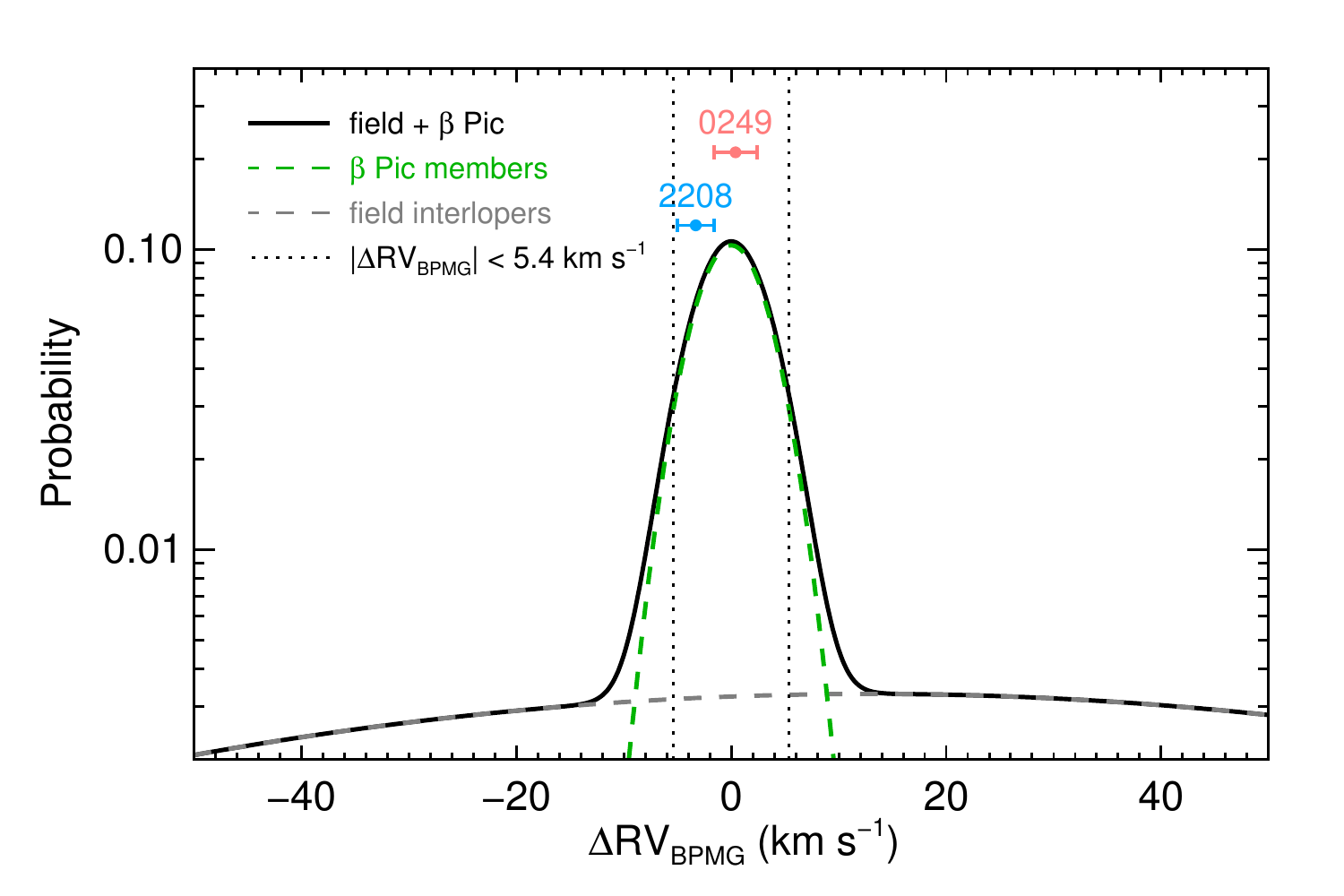}}
  \caption{\normalsize Probability distributions of
    $\Delta$RV$_{\rm BPMG}$, defined as the difference between an
    object's measured RV and the expected RV if it were a member of
    the \bpic\ moving group.  The Gaussian distributions shown here
    were derived from all objects that passed proper motion and HR
    diagram selection criteria for membership in the \bpic\ moving
    group in the ACRONYM candidate sample \citep{2017AJ....154...69S}.
    After this sample was subjected to spectroscopic follow up, 38\%
    of objects were determined to be field interlopers lacking
    evidence of youth (H$\alpha$, \ion{Li}{1}) or having RVs
    inconsistent with \bpic\ kinematics, and 62\% were confirmed as
    likely members.  The interlopers have widely varying RVs, while by
    definition the members are concentrated near
    $\Delta$RV$_{\rm BPMG}$ of zero.  Integrating the field interloper
    distribution over a range of $\Delta$RV$_{\rm BPMG}$ and dividing
    by the integral of the combined distribution (field + \bpic) over
    the same $\Delta$RV$_{\rm BPMG}$ range give the field
    contamination rate.  The original ACRONYM selection criterion of
    $|\Delta{\rm RV}_{\rm BPMG}| < 5.4$\,\kms\ gives a 4\%
    contamination rate.  Horizontal error bars are plotted at
    arbitrary probability to show the $\Delta$RV$_{\rm BPMG}$ values
    for \obj\ and \objtwo, where the error includes both the intrinsic
    dispersion in the \bpic\ group $UVW$ ellipsoid and the RV
    measurement uncertainties.  \label{fig:false-alarm}}
\end{figure}

For a given $\Delta{\rm RV}_{\rm BPMG}$ selection criterion, the
false-alarm rate is the integral of the nonmember distribution divided
by the integral of the combined distribution over the same
$\Delta{\rm RV}_{\rm BPMG}$ range (Figure~\ref{fig:false-alarm}).
Using the original ACRONYM criterion of
$|\Delta{\rm RV}_{\rm BPMG}|<5.4$\,\kms, we compute a false-alarm rate
of 4\%. Even if we consider an extremely restrictive criterion of
$|\Delta{\rm RV}_{\rm BPMG}|<0.8$\,\kms, which would let past just
\obj\ and eleven other members, the false-alarm rate would only be
reduced to 3.1\%. Therefore, contamination does not strongly depend on
the $\Delta{\rm RV}_{\rm BPMG}$ cut, as long as the cut is relatively
restrictive.

According to the binomial distribution, a false-alarm rate of 4\%
implies a 90\% probability of at least one contaminant among the 52
objects that \citet{2017AJ....154...69S} identified as members meeting
this criterion and a $<$1\% probability of $\geq$7 contaminants.
However, the appropriate sample to consider here is the set of the
twelve latest-type ACRONYM members accessible from CFHT that we have
been following up to obtain parallaxes. Among all twelve only
0.5~contaminants are expected for a 4\% false-alarm rate, and $\leq$3
should be present at 99\% confidence. \obj\ is the first object for
which we are reporting a parallax, so it is not yet possible to yet
determine if the parallaxes and improved proper motions for the other
objects are consistent with \bpic\ membership or not.

The false-alarm rate of 4\% derived for the ACRONYM sample should be
considered a conservative upper limit in the case of \obj. Firstly,
parallaxes were not available in the ACRONYM sample selection. If they
were then they would have reduced the number of interlopers that made
it to the spectroscopic follow-up stage of the ACRONYM search and
thereby reduced the nonmember component of the probability
distribution used above. Our addition of a parallax for \obj\ could
have ruled out membership by being inconsistent with the \bpic\ group
kinematic distance, but it did not. Secondly, spectroscopic binaries
that are true members may have been excluded from the ACRONYM member
sample due to orbital RV deviations. If such objects had not been
excluded that would have increased the relative number of members to
nonmembers adopted in our analysis. Thirdly, we used both young and
old stars in our analysis to improve statistics and because stars from
ACRONYM do not have homogeneous constraints on their ages. The
detection of \ion{Li}{1} absorption in \obj\ provides a much stronger
constraint than H$\alpha$ in an earlier-type ACRONYM star lying above
the lithium-depletion boundary. To determine an age constraint for
\obj, we examined other associations with measured lithium-depletion
boundaries. The components of \obj{AB} have bolometric magnitudes of
$\Mbol=11.21^{+0.24}_{-0.22}$\,mag and $11.33^{+0.24}_{-0.22}$\,mag
(Section~\ref{sec:fbol}) that are both consistent within their errors
with the boundary of $\Mbol=11.31$\,mag for $\alpha$~Persei
($85\pm10$\,Myr; \citealp{2004ApJ...614..386B}), implying a system age
$\lesssim$100\,Myr. This is somewhat stronger than the age constraint
implied by the gravity classifications of \vlg\ for both \obj{AB} and
\obj~c \citep[$\lesssim$150\,Myr; e.g.,][]{2016ApJ...833...96L}. If we
were able to restrict the field interloper population used in our
false alarm analysis to such young stars, the false-alarm probability
of 4\% would be reduced by a factor roughly equal to the number of
$>$100-Myr-old stars divided by the number of $<$100-Myr-old stars,
which is likely at least an order of magnitude.

\begin{figure} 
  \includegraphics[width=6.5in,angle=0]{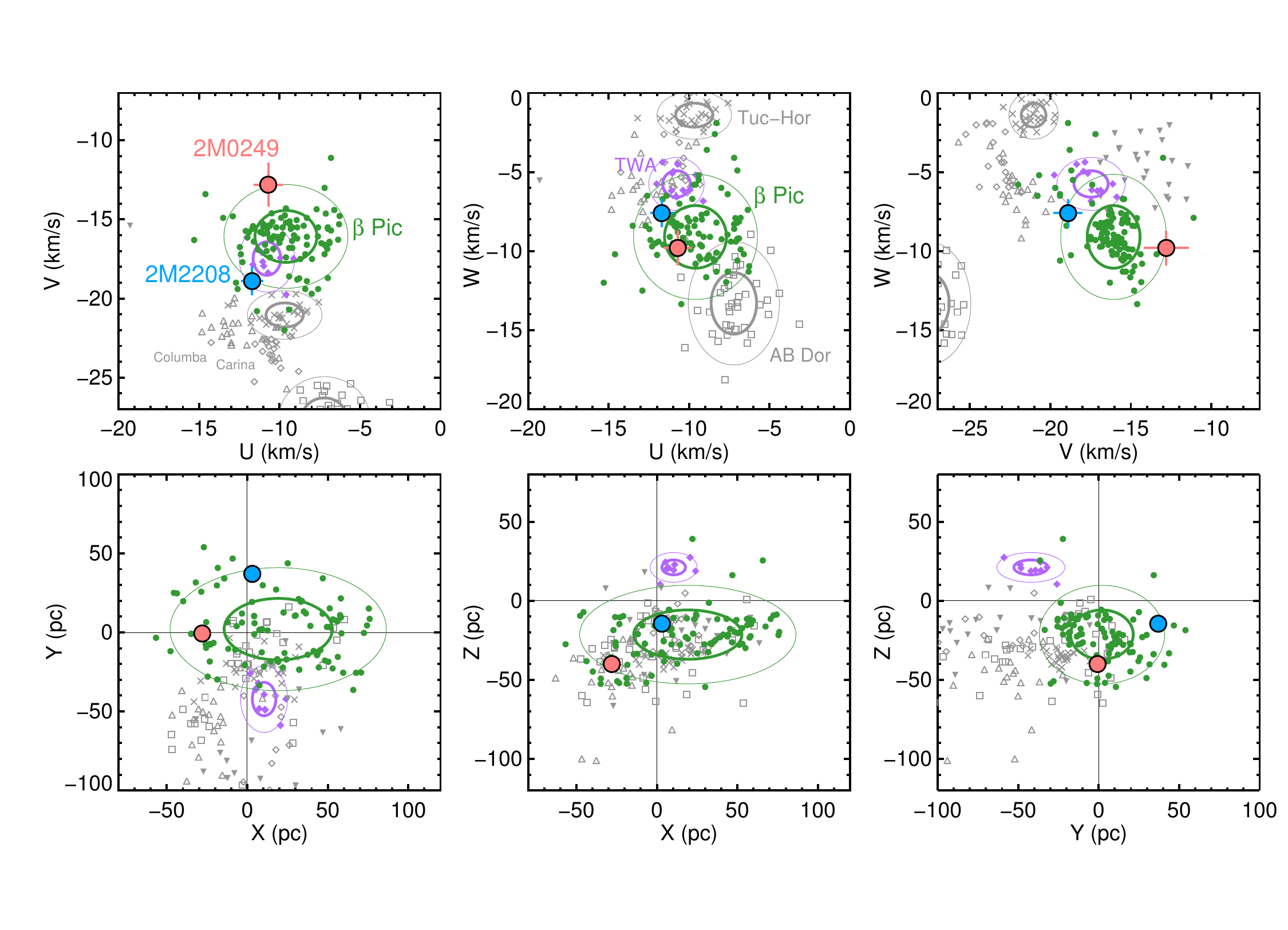}
  \vskip -0.3in
  \caption{\normalsize Kinematic ($UVW$) and spatial ($XYZ$) position
    of the \obj\ system and \objtwo\ compared to known young moving
    groups. The \bpic\ and TWA groups are highlighted in color, as the
    two nearest groups in $UVW$, but only \bpic\ also agrees in
    $XYZ$. The group members plotted here are from
    \citet{2008hsf2.book..757T}, and for \bpic\ we also include
    objects from \citet{2017AJ....154...69S}. Ellipses represent the
    1$\sigma$ and 2$\sigma$ bounds of members plotted here, and these
    are also shown for the Tuc-Hor and AB~Dor groups in $UVW$ given
    their large sample sizes, though they do not match well with \obj.
    This plot is for display purposes only. Our kinematic analysis
    uses $UVW$ ellipsoids defined by the much more restrictive, but
    spatially incomplete, lists of high-probability members compiled
    by \citet{2014MNRAS.445.2169M} and
    \citet{2018MNRAS.475.2955L}. \label{fig:uvwxyz}}
  \end{figure}

We conclude that the \obj\ system is a very likely member of the
\bpic\ moving group given its excellent kinematic agreement, low
false-alarm probability, and independent constraints on youth (lithium
and low-gravity classification) that are consistent with the age of
the group. Although we have chosen not to use the spatial $XYZ$
position of \obj\ in our membership probability analysis, because the
current census of the \bpic\ moving group has not been established to
be complete, \obj\ seems to be well within the spatial range of other
\bpic\ members (Figure~\ref{fig:uvwxyz}). It is also within the
minimum volume enclosing ellipsoid of the ``exclusive'' (smallest)
list of members shown in Figure~4 of \citet{2018MNRAS.475.2955L}.

\subsubsection{Comparison to Membership Tools\label{sec:banyan}}

Generalized tools are available in the literature that provide young
moving group membership probability estimates given input data (RA,
Dec, proper motion, parallax, and RV). We have examined membership
assessments from these tools as a point of comparison to our own
membership analysis that is tailored to the case of \obj\ and accounts
for the particular selection process of the ACRONYM search.  The
underlying assumptions used in each tool varies, such as the number of
associations included and their properties, so they can produce
wide-ranging results for the same input data. We discuss the results
from some commonly used tools before examining in detail the results
of the current version of BANYAN.

The convergent point
tool\footnote{\url{http://dr-rodriguez.github.io/CPCalc.html}} from
\citet{2013ApJ...774..101R} takes only RA, Dec, and proper motion as
inputs, and it outputs $>$10\% probabilities for all seven populations
it considers. The highest probability is 89\% for the \bpic\ moving
group, and the next highest is 57\% for Columba. This convergent point
tool also outputs a kinematic distance and RV for each group, and the
value that is closest to the system RV of \obj\ is the one for the
\bpic\ moving group (13.4\,\kms). The kinematic distance of 60\,pc for
\bpic\ membership is also consistent within the uncertainties of our
parallactic distance.
BANYAN~I\footnote{\url{http://www.astro.umontreal.ca/~malo/banyan.php}}
uses a Bayesian approach to assign an input object to one of seven
groups or the field population \citep{2013ApJ...762...88M}.  From the
proper motion, parallax, and RV of \obj\ it computes membership
probabilities of 92\% for the \bpic\ moving group and 8\% for the
field.  The convergence-style algorithm
LACEwING\footnote{\url{https://github.com/ariedel/lacewing}}
\citep{2017AJ....153...95R} considers sixteen associations, and while
the highest output probability is for the \bpic\ moving group it is
only 3\%.  None of these tools can incorporate additional information,
such as the age constraint on \obj\ of $<$100\,Myr, but they are
generally consistent with \bpic\ moving group membership.

BANYAN~$\Sigma$ is the latest version of perhaps the most widely used
membership tool, and it includes 27 young associations and the field
population \citep{2018ApJ...856...23G}. It is different from previous
versions of BANYAN \citep[e.g.,][]{2014ApJ...783..121G} in that it is
designed to deliver a uniform 90\% true-positive rate for all groups
when providing a proper motion, parallax, and RV and selecting objects
above a threshold output probability of $>$90\%.  Using v1.1 of the
BANYAN~$\Sigma$ tool \citep{jonathan_gagne_2018_1165086}, and
including both $UVW$ and $XYZ$ for \obj\ in the analysis,
BANYAN~$\Sigma$ reports probabilities of 11\% for the \bpic\ moving
group, 89\% for the field population, and negligible probabilities for
other groups.  Neither probability crosses the 90\% threshold, and
although the \bpic\ probability is low, it is not necessarily
discrepant with our analysis showing that \obj\ is a likely \bpic\
member.  This is primarily because BANYAN~$\Sigma$ does not account
for additional information about youth, such as the detection of
lithium in \obj{AB}, which should greatly reduce the prior likelihood
that the system is a field interloper.

To try to estimate the field prior used by BANYAN~$\Sigma$ at the
location of \obj, we excluded kinematic information (set proper motion
and RV errors to arbitrarily large values) and retained spatial
information (RA, Dec, and parallax). This gave a field probability of
94.8\% and Tuc-Hor probability of 4.1\%, followed by AB~Dor (0.5\%),
Columba (0.3\%), and \bpic\ (0.3\%).  The Tuc-Hor probability is
strikingly high given the location of \obj\ at
$(X,Y,Z) = (-28\pm3,-0.68\pm0.07,-40\pm4)$\,pc, a region of space
mostly devoid of known Tuc-Hor members (e.g., see Figure~10 of
\citealp{2014AJ....147..146K}).  This apparent discrepancy is likely a
consequence of the specialized way that BANYAN~$\Sigma$ determines its
priors for young moving groups relative to the field.  Normalization
factors denoted as $\ln(\alpha_k)$ are chosen to ensure a 90\%
true-positive rate for all groups, regardless of their spatial or
kinematic concentration.  This makes it difficult to quantify the
field prior relative to the \bpic\ moving group prior and how that
would change when considering only young field interlopers.
Therefore, we instead consider the true-positive and false-positive
rates reported by BANYAN~$\Sigma$.

In BANYAN~$\Sigma$, the field probability should not be interpreted as
an estimate of the false-positive rate.  For example, a hypothetical
object with proper motion, parallax, and RV giving a 90\% probability
of belonging to the \bpic\ moving group and a 10\% field probability
has a 90\% true-positive rate (by design) and a very low
false-positive rate of $1.6\times10^{-5}$ (Table~9 of
\citealp{2018ApJ...856...23G}).  This is consistent with the intention
of BANYAN~$\Sigma$ to be readily used on large input data sets, where
even such a low rate could result in hundreds of contaminants.
However, a true-positive rate of 90\% corresponds to a false-negative
rate of 10\%, which is rather conservative (i.e., equivalent to a
1.6$\sigma$ selection criterion).  Adopting such a criterion would
have the undesirable effect that any of the twelve objects in our
ACRONYM parallax follow-up sample would more likely be rejected as a
false negative (10\%) than accepted as a false positive ($\leq$4\%,
according our ACRONYM-based analysis in the previous section).  To
reduce the false-negative rate, we consider a BANYAN~$\Sigma$ \bpic\
moving group probability of 10\% (comparable to 11\% for \obj), for
which the true-positive rate is 95.57\% (i.e., equivalent to a
2.0$\sigma$ selection criterion), and the false-positive rate is still
only $2.0\times10^{-4}$ (J.\ Gagn\'e 2018, private communication). We
suggest that such a more-inclusive member selection would be
reasonable for a small sample like our parallax follow-up of twelve
ultracool dwarfs from ACRONYM.

To summarize, each of the generalized membership tools we examined
gives a higher membership probability for the \bpic\ moving group than
any other young association. The output probabilities vary widely, and
they should be considered lower limits because none of these tools can
account for the prior age information that \obj\ is young
($\lesssim$100\,Myr from lithium depletion and spectrally classified
as low gravity).  The false-positive rate predicted by BANYAN~$\Sigma$
for \obj\ is much lower than the conservative upper limit we derived
for the entire ACRONYM sample in the previous section ($\leq$4\%).
Overall, examination of these membership tools supports the conclusion
from our independent analysis, based on modeling the selection effects
of the ACRONYM sample and our CFHT parallax follow up, that the \obj\
system is a \bpic\ moving group member.

\subsection{Membership Assessment for \objtwo \label{sec:2208}}

In our previous work, \citet{2016ApJ...833...96L} identified the
L3~\vlg\ dwarf \objtwo\ as a promising candidate member of the \bpic\
moving group based on our parallax and proper motion but in need of RV
confirmation. Since then, \citet{2017ApJ...842...78V} measured an RV
of $-15.7^{+0.8}_{-0.9}$\,\kms.  Combining all these measurements
gives $(U,V,W) = (-11.7\pm0.7,-18.9\pm0.9,-7.6\pm0.9)$\,\kms, which is
3.6\,\kms\ away from the mean \bpic\ group velocity of
\citet{2014MNRAS.445.2169M} and 3.8\,\kms\ away from the mean velocity
of \citet{2018MNRAS.475.2955L}.
Projecting the \bpic\ ellipsoids into proper motion--RV space, as in
our analysis of \obj\ in the previous section, we find 3D distances
for \objtwo\ that are 2.0$\sigma$ using the
\citet{2014MNRAS.445.2169M} group parameters and 2.2$\sigma$ using
\citet{2018MNRAS.475.2955L}. (Even though \objtwo\ has similar
distances in \kms\ from the \bpic\ ellipsoids as \obj, it has slightly
larger 3D distances in $\sigma$ because its parallax, proper motion,
and RV are more precise.) The fraction of simulated \bpic\ members
that are closer than \objtwo\ in 3D space are 73\% and 82\%,
respectively. Thus, even a $>$90\% completeness criterion would
require \objtwo\ to be considered a candidate member.  Unlike \obj,
most of the 3D distance is in the RV axis, but with
$\Delta{\rm RV}_{\rm BPMG} = -3.3\pm1.7$\,\kms\ \objtwo\ is still
within $<$2$\sigma$ of the expected velocity for a member and would
easily pass the RV criterion used for the ACRONYM sample
(Figure~\ref{fig:false-alarm}).  Therefore, based on kinematics alone,
\objtwo\ appears to be a likely member of the \bpic\ moving group.

Spatially, \objtwo\ coincides with other published members of the
\bpic\ moving group given its position of
$(X,Y,Z) = (3.1\pm0.2,37.0\pm2.4,-14.5\pm0.9)$\,pc
(Figure~\ref{fig:uvwxyz}).  Compared only to stars belonging to the
most restrictive member lists, it is somewhat discrepant in the $Y$
axis; for example, it lies just outside the minimum volume enclosing
ellipsoid of \citet{2018MNRAS.475.2955L}.  This may not be a major
cause for concern, as it has been suggested that the spatial
distribution of members may be larger than is currently known,
especially for widely dispersed groups like \bpic\ and AB~Dor
\citep[e.g.,][]{2016ApJ...833...96L, 2017AJ....153...18B,
  2018ApJ...852...55D}.

BANYAN~$\Sigma$ reports a surprisingly low probability of 0.8\% for
\bpic\ membership (99.2\% for field) for \objtwo. This in contrast
with the previous \bpic\ probability of 18\% computed by
\citet{2016ApJ...833...96L} using BANYAN~II
\citep{2014ApJ...783..121G} with the same proper motion and parallax,
as well as a membership probability of 96.5\% using BANYAN~I
\citep{2013ApJ...762...88M} with both a parallax and RV.  The only
change in the observations since \citet{2016ApJ...833...96L} is the
addition of an RV from \citet{2017ApJ...842...78V}, which is
consistent with the expected value for \bpic\ within 2$\sigma$ even
according to BANYAN~$\Sigma$ (optimal RV of $-13.6\pm0.7$\,\kms).  As
for other membership tools, they also give a higher membership
probability for the \bpic\ moving group than any other young
association, with 94\% from the convergent point tool
\citep{2013ApJ...774..101R} and 19\% from LACEwING
\citep{2017AJ....153...95R}.

The discrepancy between \objtwo\ passing the same kinematic selection
criteria as \obj\ and the $>$10$\times$ lower membership probability
from BANYAN~$\Sigma$ may be related to the fact that BANYAN~$\Sigma$
does not account for evidence of youth that would reduce the fraction
of field interlopers.  Given the \vlg\ gravity classification,
\objtwo\ is quite young. Among parallax-confirmed members of young
moving groups, \citet{2016ApJ...833...96L} found that the \vlg\
classification becomes less prevalent by an age of $\approx$150\,Myr
compared to the intermediate-gravity classification \intg. This trend
is less clear among (less definitive) candidate lists for moving
groups using objects that lack parallaxes or RVs (e.g.,
\citealp{2015ApJS..219...33G, 2016ApJS..225...10F}).  Still, all
previously known parallax- and RV-confirmed ultracool dwarf members of
the \bpic\ moving group are classified as \vlg.  Therefore, \objtwo\ is
independently known to be consistent with the age of the \bpic\ group,
which reduces the probability that it is a field interloper.

We conclude that \objtwo\ is a possible member of the \bpic\ moving
group, but given the discord with BANYAN~$\Sigma$, we consider its
status ambiguous.  Unlike \obj, there is little room for improving the
observations of \objtwo\ (system RV, distance, proper motion), so a
more robust look at its membership will need a more complete census of
\bpic\ members.  \Gaia\ will not map the spatial distribution of young
L~dwarfs out to the necessary distances for such work, due to their
faintness (e.g., \objtwo\ itself does not have an entry in \Gaia~DR2).
However, it should be possible to use the higher mass M~dwarfs to
better determine the spatial distribution of the \bpic\ moving group.

\subsection{Color--Magnitude Diagram \label{sec:cmd}}

Combining our photometry with the distance modulus derived from our
parallax ($m-M = 3.44^{+0.21}_{-0.23}$\,mag) allows us to compute
absolute magnitudes and compare them to the polynomial relations of
magnitude versus spectral type from \citet{2016ApJ...833...96L}.  For
a spectral type of L$2\pm1$~\vlg, the polynomials give
$M_J=12.4\pm1.0$\,mag and $M_K=10.6\pm0.9$\,mag, where the
uncertainties are a quadrature sum of the rms of objects used in the
fit about the polynomial curve ($\pm$0.6\,mag and $\pm$0.4\,mag,
respectively) and the propagation of the spectral type uncertainty
($\pm$0.8\,mag).  The absolute magnitudes of \obj~c are
$M_J=13.20^{+0.22}_{-0.24}$\,mag and $M_K=11.34^{+0.22}_{-0.24}$\,mag,
which are 0.8$\sigma$ and 0.9$\sigma$ fainter than the polynomial and
thus consistent with being a normal object for its spectral type and
gravity classification.  Its faintness also suggests that \obj~c is
not likely to be an unresolved, near-equal-flux binary.  For the
M6~\vlg\ integrated-light spectral type of \obj{AB}, the
\citet{2016ApJ...833...96L} polynomial gives $M_K=7.2\pm0.4$\,mag.
(The polynomial is only valid for types of M6 and later, so we cannot
reliably estimate the additional error due to a $\pm$1~subtype
uncertainty.) Our resolved $K$-band magnitudes of the primary
($M_K=8.29^{+0.21}_{-0.23}$\,mag) and secondary
($M_K=8.41^{+0.21}_{-0.23}$\,mag) are 1.1\,mag and 1.2\,mag fainter,
respectively, than the polynomial but again not overly discrepant
within the scatter about the polynomials. For context, there are other
examples of low-gravity M6 dwarfs with similar or fainter absolute
magnitudes, such as HD~1160B ($M_K = 8.83\pm0.16$\,mag;
\citealp{2012ApJ...750...53N}).

Figure~\ref{fig:cmd-bpmg} shows the components of the \obj\ triple
system alongside members of the \bpic\ moving group on an IR
color--magnitude diagram.  \obj{AB} has a color similar to other
late-M members of \bpic, like PZ~Tel~B (M7) and 2MASS~J0335+2342
(M7~\vlg), and each component has a comparable or somewhat brighter
absolute magnitude, if we assume the two components have similar
infrared colors.  Likewise, \obj~c lies near the other \bpic\ objects
with L1--L3 spectral types: \bpic~b and \objtwo.

\subsection{Estimated Masses \label{sec:mass}}

In order to derive physical properties, we rely on the predicted
luminosity (\Lbol) as a function of mass and age from evolutionary
models.
We compute luminosities by combining the bolometric fluxes derived in
Section~\ref{sec:fbol} with our parallax measurement of
$20.5\pm2.1$\,mas. For \obj~c we find
$\log(\Lbol/\Lsun) = -4.00\pm0.09$\,dex. For \obj{AB} we divide its
integrated-light luminosity assuming that the nearly equal-flux
components ($\Delta{K} = 0.123\pm0.005$\,mag) have negligible
difference in their $K$-band bolometric corrections compared to the
uncertainty in the distance modulus ($3.44^{+0.21}_{-0.23}$\,mag).
Thus, adopting our $K$-band flux ratio as the bolometric flux ratio
results in component luminosities of
$\log(\Lbol/\Lsun) = -2.59\pm0.09$\,dex and $-2.64\pm0.09$\,dex.

No single model grid completely covers the luminosities of all
three components, so we use the \citet{2015A&A...577A..42B} tracks for
\obj{AB} and \citet{2008ApJ...689.1327S} hybrid tracks for \obj~c. In
a similar fashion as \citet{2017ApJS..231...15D}, we use Monte Carlo
rejection sampling with uniformly distributed masses and ages as the
initial input. For each trial mass and age, we compute
$$\chi^2 = \left(\frac{\log(L_{\rm bol, model})-\log(L_{\rm bol})}{\sigma_{\log(L_{\rm bol})}}\right)^2 + \left(\frac{t-{\rm 22\,Myr}}{\rm 6\,Myr}\right)^2,$$
from which we compute a rejection probability
$p = e^{-(\chi^2-\min(\chi^2))/2}$.  We then draw random, uniformly
distributed variates $u$ and reject samples where $p < u$. This method
allows us to properly account for the possibility that objects of
different masses and ages have the same luminosity due to deuterium
fusion, which can be important at such young ages. The age prior of
$22\pm6$\,Myr is based on the most recent measurement of the lithium-depletion boundary in the \bpic\ moving group from
\citet{2017AJ....154...69S}, which is consistent with the previous
lithium-depletion age of $23\pm3$\,Myr from
\citet{2014MNRAS.438L..11B} and the isochronal age of $24\pm3$\,Myr
from \citet{2015MNRAS.454..593B}. 

Table~\ref{tbl:prop} gives the resulting masses of the three
components. As expected given its M6~\vlg\ spectral type, the tight
binary \obj{AB} is estimated to be a pair of brown dwarfs, while the
L2~\vlg\ companion's mass of $11.6^{+1.3}_{-1.0}$\,\Mjup\ is likely
below the deuterium fusion boundary \citep[$\approx$13\,\Mjup;
e.g.,][]{2011ApJ...727...57S}. For comparison, we also derived masses
for other late-type members of \bpic\ using the
\citet{2008ApJ...689.1327S} models and the same rejection sampling
method.
For the free-floating objects, we computed our own bolometric fluxes
using the same method described in Section~\ref{sec:fbol}, except that
in these cases we simply used published IRTF/SpeX spectra combined
with BT-Settl models at other wavelengths. This included
2MASS~J0335+2342, SDSS~J0443+0002, 2MASS~J1935$-$2846,
2MASS~J2013$-$2806, PSO~J318.5$-$22, and \objtwo.
For companions, we used published values from the literature for
\bpic~b and 51~Eri~b, and \Ks-band photometry combined with the
spectral type--BC$_{\Ks}$ relation for young objects from
\citet{2015ApJ...810..158F} for PZ~Tel~B and HR~7329B.  All the
luminosities and derived masses are given in Table~\ref{tbl:bpic}.
We emphasize that none of our quoted mass errors attempt to account
for unknown systematic uncertainties in the evolutionary models, which
are likely larger than the random errors in cases where luminosity is
measured very precisely. We also note that formal mass errors are
larger for the higher mass components \obj{AB} than for the wide
companion \obj~c because the evolutionary models predict that such
young, massive brown dwarfs have similar masses at a fixed age. In
other words, isomass tracks pile up on an \Lbol--age diagram due to
deuterium fusion; for example, see Figure~1 in \citet{2001RvMP...73..719B}.

\subsection{\Gaia~DR2 \label{sec:gaia}}

A parallax and proper motion for \obj{AB} (in integrated light) are
available from \Gaia~DR2 \citep{2016A&A...595A...1G,
  2018arXiv180409365G}.  Binary orbital motion can impact the parallax
and proper motion in an unpredictable, systematic way, especially when
there are relatively few independent observation epochs (DR2 reports
{\tt visibility\_periods\_used = 10} for this system). Indeed, the
excess source noise ($\epsilon_i$) reported in DR2 for \obj{AB} is
0.70\,mas at a significance of 37$\sigma$, suggesting that systematic,
correlated noise from orbital motion impacts the five-parameter DR2
solution.  There is evidence from hierarchical M~dwarf triple systems,
where one component is single and the other is unresolved in \Gaia\
(e.g., GJ~1245 and GJ~2069), that parallax systematics for unresolved
binaries can be up to at least 2\,mas ($\approx20\sigma$) with even
larger proper motion systematics.  Therefore, we have chosen not to
use the DR2 astrometry for \obj{AB} in our analysis until the accuracy
of DR2 parallaxes for close binaries like this can be more carefully
vetted.  However, we briefly consider here what the difference in our
analysis would be if we used the DR2 parallax and proper motion.

The \Gaia~DR2 parallax of $15.11\pm0.10$\,mas is 2.6$\sigma$ lower
than our value for \obj{AB} and 1.4$\sigma$ lower than for \obj~c.
The \Gaia~DR2 proper motions are 1.2$\sigma$ higher in RA and
2.0$\sigma$ lower in Dec.  Using the \Gaia~DR2 parallax and proper
motion with the RV from \citet{2017AJ....154...69S} gives
$(U,V,W) = (-11.4\pm0.8,-18.81\pm0.14,-9.3\pm1.1)$\,\kms\ and
$(X,Y,Z) = (-38.2\pm0.3,-0.924\pm0.006,-54.1\pm0.4)$\,pc. The most
significant difference with our kinematics is in $V$, which our CFHT
astrometry shows is 3\,\kms\ higher than the mean \bpic\ group motion
but \Gaia~DR2 indicates is 3\,\kms\ lower. Rerunning our kinematic
analysis using \Gaia~DR2 gives a 3D distance from \bpic\ in proper
motion--RV space of 2.0$\sigma$, still closer to the mean than 25\% of
simulated members.  Our false alarm analysis is unchanged because it
is based on the selection criteria of the ACRONYM search, which
\obj{AB} would still pass.

We also note that the smaller \Gaia~DR2 parallax implies brighter
absolute magnitudes for all three components.  For \obj{AB} this would
mean bolometric magnitudes of $\Mbol = 10.51\pm0.03$\,mag and
$10.63\pm0.03$\,mag, which in turn would imply a younger age upper
limit from the detection of lithium.  \obj{AB} would be younger than
$\alpha$~Persei ($85\pm10$\,Myr), where the lithium-depletion boundary
is at $\Mbol = 11.31\pm0.15$\,mag.  It would instead be consistent
with the next youngest measured lithium-depletion boundary in IC~2391
(45\,Myr, $\Mbol = 10.24\pm0.15$\,mag; \citealp{2004ApJ...614..386B}).
An even younger upper limit on the age would make it correspondingly
less likely that the \obj\ system is a field contaminant rather than a
member of the very young \bpic\ moving group ($22\pm6$\,Myr).  We
therefore conclude that the \Gaia~DR2 results are generally consistent
with our membership assessment based on our CFHT astrometry.  The
0.66\,mag brighter $J$- and $K$-band absolute magnitudes would also
bring all three components of \obj{ABc} closer to the
\citet{2016ApJ...833...96L} polynomial relations. \obj~c would also
lie much closer to \bpic~b on the color--magnitude diagram.

The DR2 parallax also implies higher luminosities.  The components of
\obj{AB} would have $\log(\Lbol/\Lsun) = -2.32\pm0.02$\,dex and
$-2.37\pm0.02$\,dex, which would make them more luminous than other
known ultracool dwarfs in the \bpic\ moving group but still normal for
a spectral type of M6.  Their estimated masses would be somewhat
higher, at $75^{+12}_{-11}$\,\Mjup\ and $69^{+13}_{-9}$\,\Mjup, but
still consistent with being brown dwarfs.  \obj~c would have
$\log(\Lbol/\Lsun) = -3.73\pm0.02$\,dex, i.e., 0.05\,dex (1.4$\sigma$)
more luminous than \bpic~b, and a mass of $13.2^{+0.4}_{-0.2}$\,\Mjup\
(1.2$\sigma$ higher than the mass derived using our CFHT parallax).

\section{Discussion}

\subsection{Implications for the \bpic\ Moving Group}

The \obj{AB}~c system increases the total number of ultracool
($\geq$M6) members of the \bpic\ moving group from seven to ten.
Table~\ref{tbl:bpic} summarizes these members, as well as other
possible members that are lacking either parallax or RV for
confirmation, and also those based on proper motion alone.
\Gaia\ will soon enable a reassessment of all these objects, either
directly via high-precision parallaxes and proper motions for the
brightest ones or indirectly via an improved census of local moving
groups.  In the meantime, it is noteworthy that current methods of
assessing group membership can still disagree significantly. The \obj\
system is one such example, as it is not classified as a
high-probability ($P\geq90$\%) \bpic\ member from BANYAN~$\Sigma$,
even though we reaffirm its membership as originally determined by
\citet{2017AJ....154...69S}.
\objtwo\ is a more puzzling case of an object with kinematics that
would make it a likely \bpic\ member according to our analysis but
with widely varying probabilities from generalized membership tools,
as low as 0.8\% from BANYAN~$\Sigma$ \citep{2018ApJ...856...23G}, so
we must conclude that its membership is ambiguous.

\obj{AB}c is the first ultracool triple system found in the \bpic\
moving group.  The only other \bpic\ system containing more than one
late-type component is the possible member DENIS~J0041$-$5621AB
\citep[integrated-light type M7.5~\vlg;][]{2015ApJS..219...33G,
  2017AJ....154...69S}, a 7\,AU binary with an estimated orbital
period of 126\,yr \citep{2010A&A...513L...9R}.  To our knowledge, no
other ultracool triple systems are known in any other young moving
groups.  The only known binaries are 2MASS~J1119$-$1137AB
\citep{2017ApJ...843L...4B}, which is a likely TWA member, and
DENIS~J0357$-$4417AB \citep{2003AJ....126.1526B}, which is a candidate
Tuc-Hor member \citep{2014ApJ...783..121G, 2015ApJ...798...73G}.
\obj{AB}c is the sixth known substellar triple system, i.e., composed
entirely of likely brown dwarfs \citep{2005AJ....129..511B,
  2013ApJ...778...36R, 2016ApJ...818L..12S, 2017ApJS..231...15D}.
Compared to the 1950\,AU separation for \obj~c, the other known
substellar triples are more compact, with the widest of them
(VHS~J1256$-$1257; \citealp{2016ApJ...818L..12S}) having an outer pair
separation of 100\,AU, and the rest having outer separations of
2--27\,AU.

In contrast to other known young ultracool binaries, \obj{AB} is much
tighter (projected separation $2.17\pm0.22$\,AU at discovery). Its
estimated orbital period of $\approx$8\,yr makes it likely to yield
the first dynamical mass measurement in the \bpic\ moving group in the
substellar regime.  In addition to the usual strong tests of
substellar models enabled by dynamical masses
\citep[e.g.,][]{2008ApJ...689..436L, 2010ApJ...721.1725D,
  2016ApJ...827...23D, 2012ApJ...751...97C}, this binary will yield
the first substellar cooling age (i.e., using luminosity and mass) for
a young moving group. Thus, \obj{AB} will enable a unique
cross-calibration of substellar evolutionary model tracks by comparing
to ages from the lithium-depletion boundary and stellar isochrone
methods.  The cooling rate of brown dwarfs predicted by evolutionary
models has only been independently tested where brown dwarf binaries
orbit young stars with gyrochronology-derived ages
\citep{2009ApJ...692..729D, 2014ApJ...790..133D} or where a brown
dwarf orbits an older star with a (less precise) isochronal or
kinematic age \citep[e.g.,][]{2008ApJ...678..463I,
  2018AJ....155..159B}.  Tests of substellar evolutionary models are
especially needed at the young age of \bpic\ as they are frequently
used to infer the physical properties of planetary-mass companions.

\subsection{An Unusual System Architecture}

\obj~c ($11.6^{+1.3}_{-1.0}$\,\Mjup) is unique among companions at or
below the deuterium-fusion boundary given its wide separation
($1950\pm200$\,AU) and the fact that it orbits a very low-mass binary
($48^{+13}_{-12}$\,\Mjup\ and $44^{+14}_{-11}$\,\Mjup).
Figure~\ref{fig:sep-q} shows the mass ratios of all known directly
imaged planetary-mass companions ($\lesssim$13\,\Mjup) as a function
of their projected separation.
There are only five other companions with similarly wide separations
($\gtrsim$10$^3$~AU): the AB~Dor member GU~Psc~b
\citep[$11\pm2$\,\Mjup\ at 2000~AU;][]{2014ApJ...787....5N}, the
Ophiuchus member SR~12~c \citep[$13\pm2$\,\Mjup\ at
1100~AU;][]{2011AJ....141..119K}, the young field objects Ross~458~c
\citep[$9\pm3$\,\Mjup\ at 1190~AU;][]{2010MNRAS.405.1140G} and
TYC~9486-927-1B \citep[12--15\,\Mjup\ at
6900~AU;][]{2016MNRAS.457.3191D}, and the old field object
WD~0806$-$661~b \citep[$7.5\pm1.5$\,\Mjup\ at
6900~AU;][]{2012ApJ...744..135L}.\footnote{In order to quote system
  properties consistently we use parameters given in Table~1 of the
  review by \citet{2016PASP..128j2001B} when available.}  These host
stars range from 0.3--2\,\Msun\ (adopting the progenitor mass for
WD~0806$-$661) and thus represent companion mass ratios of $\sim$0.03
or much lower, in contrast to the $\sim$0.1 mass ratio of \obj~c.  (We
adopt the combined mass of \obj{AB} as the host mass for the system.)

Among planetary-mass companions at all separations, few have hosts
with such low masses as \obj{AB}, even using its combined mass
(50--150\,\Mjup\ at 2$\sigma$). The two clearest examples are
2MASS~J1207$-$3932~b ($5\pm2$\,\Mjup) that orbits a 25\,\Mjup\ TWA
member \citep{2004A&A...425L..29C} and 2MASS~J0441+2301Bb
($10\pm2$\,\Mjup), which orbits the $19\pm3$\,\Mjup\ tertiary component
of a quadruple system in Taurus \citep{2010ApJ...714L..84T,
  2011ApJ...731....8K,
  2015ApJ...811L..30B}.\footnote{2MASS~J0441+2301Bab is itself a wide
  companion (1800~AU) to a pair of $200^{+100}_{-50}$\,\Mjup\ and
  $35\pm5$\,\Mjup\ objects.}  The slightly higher-mass companions
FU~Tau~B \citep[$\approx$16\,\Mjup;][]{2009ApJ...691.1265L} and
2MASS~J0219$-$3925B \citep[$14\pm1$\,\Mjup;][]{2015ApJ...806..254A}
orbit a 50\,\Mjup\ brown dwarf in Taurus and a 110\,\Mjup\ star in
Tuc-Hor, respectively.\footnote{VHS~J1256$-$1257~b was originally
  identified as an $11^{+10}_{-2}$\,\Mjup\ companion to a pair of
  65\,\Mjup\ objects \citep{2015ApJ...804...96G}, but
  \citet{2016ApJ...818L..12S} noted the published parallax may have
  underestimated systematic errors and derived component masses of 73,
  73, and 35\,\Mjup\ from a spectrophotometric distance, so we exclude
  it here.}  These systems' mass ratios range from 0.13--0.5,
comparable to but somewhat higher than \obj~c.
In addition, there are a number of potentially planetary-mass brown
dwarfs on close-in orbits of other brown dwarfs with similar or only
slightly higher masses that resemble scaled-down binary star systems:
SDSS~J2249+0044AB \citep[L3+L5;][]{2010ApJ...715..561A},
CFBDSIR~J1458+1013AB \citep[T9+Y;][]{2011ApJ...740..108L},
WISE~J1217+1626AB \citep[T9+Y0;][]{2012ApJ...758...57L},
WISE~J0146+4234AB \citep[T9+Y0;][]{2015ApJ...803..102D}, and
2MASS~J1119$-$1137AB \citep[L7+L7;][]{2017ApJ...843L...4B}.
In short, \obj~c is the only planetary-mass companion with both a very
wide separation ($>10^3$~AU) and relatively high mass ratio
($M_{\rm comp}/M_{\rm host}\gtrsim0.1$), suggesting that it is more
binary-like than planet-like.

\subsection{Formation Scenarios}

The mass ratio of \obj~c to its host binary is consistent with typical
stellar triple systems \citep[e.g.,][]{2017ApJS..230...15M}. But even
viewed as a very low-mass analog of stellar systems, \obj~c is still
unusual for the large separation of its tertiary orbit. At a projected
separation of 1950\,AU, it is only weakly bound to \obj{AB}.  Although
theoretical work suggests that such wide systems can form via the
dissolution of the parent cluster \citep{2010MNRAS.404.1835K}, this
route is less likely at low component masses, and the progenitor
\bpic\ cluster may never have been dense enough to facilitate
capture. Alternatively, turbulent fragmentation models of star
formation do predict that objects can form at wide separations
\citep[e.g.,][]{2009ApJ...703..131O, 2012MNRAS.419.3115B}.  In this
scenario, the \obj\ system would represent the low-mass tail of the
star formation process, drawn from an initial mass function that
sharply drops toward very low masses \citep{2003PASP..115..763C}. This
is consistent with previous surveys for wide-orbit planetary-mass
companions that find such systems are rare in young moving groups
\citep[e.g.,][]{2016ApJ...821..120A, 2017AJ....154..129N}.

An alternative hypothesis for the origin of \obj~c is that it formed
in a disk around the binary brown dwarf pair and was scattered outward
via dynamical interactions. For the masses and separations involved,
this scenario is disfavored for several reasons. First, formation via
the bottom-up core accretion process is strongly disfavored based on
simple mass requirements. Given the combined mass of the brown dwarf
binary host ($\sim$100\,\Mjup), even a disk with a total gas mass
equal to the central masses would contain only $\sim$1\,\Mjup\ of
solids.  This mass is insufficient to trigger runaway gas accretion,
even under favorable conditions \citep{1986Icar...67..391B,
  2014ApJ...786...21P}.  Formation of a tertiary in the disk by
gravitational instability instead would also require very high disk
masses, which have yet to be observed around brown dwarfs
\citep[e.g.,][]{2016A&A...593A.111T}.  To achieve the current system
architecture in this scenario, one must also invoke dynamical
interactions between the three objects.  While binaries are efficient
ejectors of planetary-mass objects \citep{2016MNRAS.461.1288S}, the
architecture of the \obj\ system is somewhat disfavored based on
energetic arguments.  The Keplerian velocity of the tertiary is
roughly 3\% of the Keplerian velocity of the host binary. Typical
scattering encounters would send the tertiary outward at velocities
10$\times$ higher than this \citep{2006tbp..book.....V}. Fine-tuning
of the interaction would be required to achieve something so
marginally bound.

In both scenarios, the fact that the PA of the tertiary companion
($228\fdg649\pm0\fdg013$) is very close to that of the inner binary
($233\fdg1\pm0\fdg3$) is most likely a coincidence.  Orbit monitoring
of the inner binary is needed to determine the actual PA of the
orbital node of \obj{AB}, but even if it is aligned with the companion
PA it would be difficult to physically explain orbital alignment over
three orders of magnitude in separation.

\subsection{A Control Sample for Studying Giant Planet Formation}

\begin{figure} 
  \vskip -0.3in
  \centerline{\includegraphics[width=4.5in,angle=0]{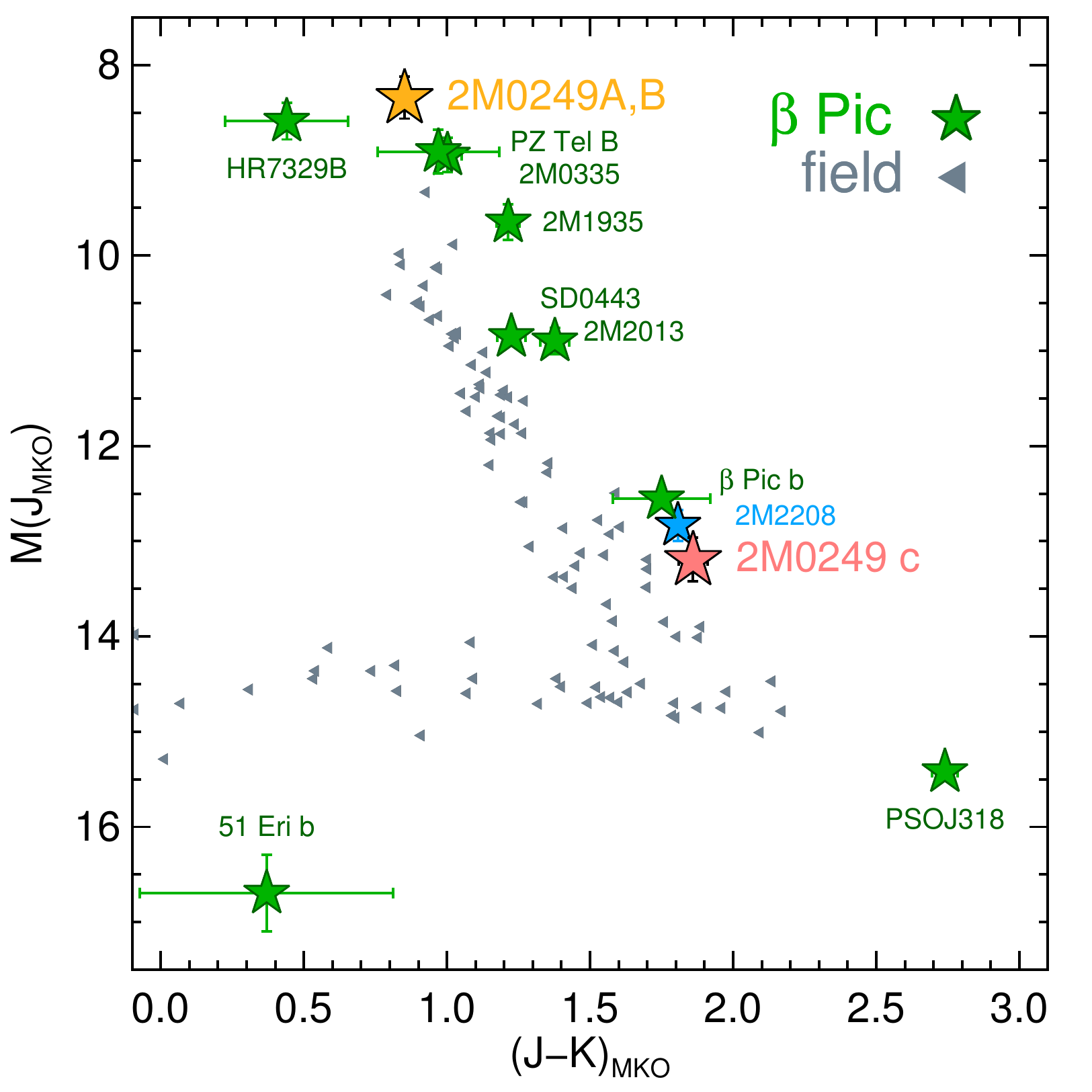}}
  \caption{\normalsize Color--magnitude diagram showing the \obj\
    system, the possible member 2MASS~J2208+2921, other ultracool
    members of the \bpic\ moving group with parallaxes, and field
    ultracool dwarfs.  Our new companion \obj~c (L2~\vlg) lies in a
    similar part of the diagram as the planet \bpic~b (L2) and the
    free-floating object 2MASS~J2208+2921 (L3~\vlg).  The components
    of \obj{AB} (M6~\vlg) have similar colors and magnitudes to the
    companions HR~7329B (M7.5) and PZ~Tel~B (M7) as well as the
    free-floating object 2MASSJ~0335+2342 (M7~\vlg). For \obj{AB} we
    show the integrated-light photometry divided by two, i.e.,
    assuming equal fluxes and colors.  [Field dwarfs are from the
    Database of Ultracool Parallaxes at
    \url{http://www.as.utexas.edu/~tdupuy/plx/}
    \citep{2012ApJS..201...19D}.] \label{fig:cmd-bpmg}}

\end{figure}

\begin{figure} 
  \vskip -0.3in
  \centerline{\includegraphics[width=6.0in,angle=0]{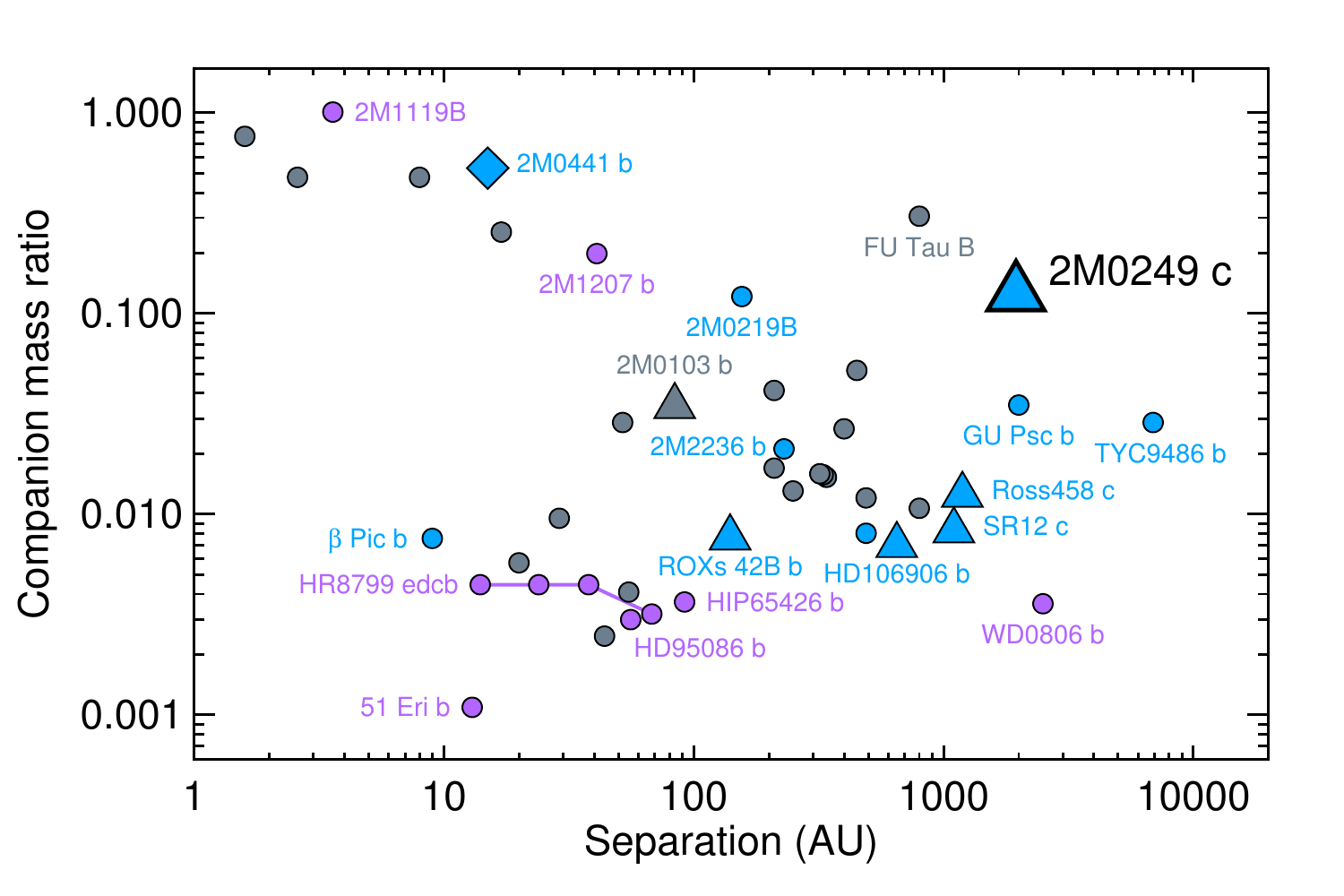}}
  \caption{\normalsize Companion mass ratio
    ($M_{\rm comp}/M_{\rm host}$) as a function of separation for
    directly imaged companions that have mass estimates near or below
    the deuterium-fusion limit (13\,\Mjup).  Symbol shapes indicate
    companions to single stars (circles), binaries (triangles), or a
    member of a quadruple system (diamond).  Symbol colors correspond
    to estimated masses from hot-start models: purple for the lowest
    mass objects ($\leq$10\,\Mjup\ even including 1$\sigma$
    uncertainties); blue for slightly higher mass objects
    ($\leq$15\,\Mjup\ even including 1$\sigma$ uncertainties); and
    gray for all other objects ($>$15\,\Mjup). \obj~c
    ($11.6^{+1.3}_{-1.0}$\,\Mjup) has an unusual combination of high
    mass ratio ($\sim$0.1) and wide separation ($1950\pm200$\,AU),
    strikingly different from the other planetary-mass companions in
    the \bpic\ moving group (\bpic~b and 51~Eri~b), which are among
    the smallest separation, lowest mass ratio companions known.  For
    this plot, we used the compilation of system properties from
    \citet{2016PASP..128j2001B} and added HIP~65426~b
    \citep{2017A&A...605L...9C}, 2MASS~J2236+4751~b
    \citep{2017AJ....153...18B}, 2MASS~J1119$-$1137AB
    \citep{2017ApJ...843L...4B}, and HD~203030B
    \citep{2006ApJ...651.1166M, 2017AJ....154..262M}.  When a
    companion orbits a binary, we use the total mass of the binary to
    compute the mass ratio.  \obj~c would thus have a plotted mass
    ratio $\approx$2$\times$ higher if we used the primary component's
    mass ($48^{+13}_{-12}$\,\Mjup) instead. (The blue circle without a
    label near HD~106906~b is HD~203030B.) \label{fig:sep-q}}

\end{figure}

\obj~c is the first wide-orbit companion ($\gtrsim$10$^3$\,AU) to have
properties so similar to a close-in planet from the same moving group,
in this case \bpic~b (9\,AU).  The existence of a third nearly
identical, but free-floating, possible member of the \bpic\ group
\objtwo\ would make for a unique trio of planetary-mass objects.
There are other well-known analogs; for example, the HR~8799 planets
have spectra, colors, and magnitudes similar to that of free-floating
objects like PSO~J318.5338$-$22.8603 \citep{2013ApJ...777L..20L} and
WISEP~J004701.06+680352.1 \citep{2015ApJ...799..203G}, but no such
objects are kinematically associated with the HR~8799 system.
\obj~c (and possibly \objtwo) are therefore ``twins,'' not merely
analogs, of \bpic~b because they all formed from the same natal
material.  Figure~\ref{fig:cmd-bpmg} illustrates that the colors and
magnitudes of these three objects are comparable within the
uncertainties, as expected for having similar spectral types and the
same age.  Similarly, Table~\ref{tbl:bpic} shows that they have
estimated masses that are consistent within the uncertainties.

If different formation mechanisms produced these objects, then their
spectra could contain evidence of their divergent pasts.  As noted
above, we suspect that \obj~c arose from a star-formation-like process
of global, top-down gravitational collapse in the same way as the
free-floating object \objtwo.  On the other hand, \bpic~b bears
architectural resemblance to planetary systems and thus may have
formed via core accretion. Core accretion models and observations of
solar system gas giants show substantial metal enrichment
\citep[e.g.,][]{1982AREPS..10..257S, 2017Sci...356..821B}.  Thus, if
\bpic~b is a scaled-up gas giant ($\approx$13\,\Mjup), then we may
expect to see substantial metal enrichment in its atmosphere.
\citet{2016ApJ...831...64T} have shown that transiting planets over a
wide range of masses ($\sim$0.1--10\,\Mjup) have enhanced metal
content with respect to their host stars, with
$Z_{\rm pl}/Z_{\star} \approx 10\times(M_{\rm pl}/\Mjup)^{-0.5}$.
While this correlation was derived from bulk density measurements, the
amount of heavy elements is so large that it implies a significant
amount of the metals are likely present in planetary atmospheres as
well as their cores.

Some have proposed that planets like \bpic~b could form via
gravitational instability in a disk
\citep[e.g.,][]{2011ApJ...731...74B}, though most models suggest that
it is unlikely \citep[e.g.,][]{2010ApJ...710.1375K,
  2013ApJ...772L..15R}.  In principle, metallicity enhancement is also
possible in this case \citep{2014prpl.conf..643H}, either from dust
trapping in spiral arms \citep{2009MNRAS.398L...6C} or from accretion
of dust and planetesimals \citep{2010ApJ...724..618B}.  However, in
the modern paradigm in which most planetesimals are formed via the
streaming instability \citep{2005ApJ...620..459Y}, such enhancement
may be suppressed.  Thus, if \bpic~b showed metal enhancement compared
to \obj~c, this could be a convincing signature of core accretion
operating at very high planetary masses.  Measuring elemental
abundances via molecules in ultracool atmospheres is challenging, but
significant progress has already been made on a number of directly
imaged planets \citep[e.g.,][]{2013Sci...339.1398K,
  2015ApJ...804...61B, 2016ApJ...817..166S, 2017AJ....154...91L}.  The
very wide separation of \obj~c (40\arcsec) and its nearly equatorial
declination will make it amenable to such follow-up observations from
nearly any ground-based telescope without needing high-contrast AO.

Formation may also affect the typical rotation rates of planetary-mass
objects.  The relatively slow rotation of solar system planets is a
well-known problem requiring some mechanism to shed angular momentum
\citep[e.g.,][]{1996Icar..123..404T}.  The equatorial velocity of
\bpic~b was measured to be 25\,\kms\ by \citet{2014Natur.509...63S},
consistent with an extrapolation of the trend among solar system
planets for faster rotation to higher masses.  The free-floating
\bpic\ moving group member PSO~J318.5$-$22
($6.5^{+1.2}_{-0.8}$\,\Mjup) is lower in mass than \bpic~b and shows a
slower equatorial velocity ($17.5\pm1.5$\,\kms), as would be expected
if it followed the same rotation--mass relation
\citep{2016ApJ...819..133A}.  The results of
\citet{2016ApJ...818..176Z} and \citet{2018NatAs...2..138B} are also
consistent with a single relationship between companions and
free-floating objects at planetary masses.  However, studies to date
have been unable to hold both mass \emph{and} age constant when
testing for differences between the rotation of free-floating objects
and companions.  Fortunately, \objtwo\ already has a published
rotation period of $3.5\pm0.2$\,hr \citep{2015ApJ...799..154M} and
$\vsini=40.6^{+1.3}_{-1.4}$\,\kms\ \citep{2017ApJ...842...78V}, both
of which imply significantly more rapid rotation than \bpic~b.  This
is suggestive of the split in behavior that is expected from the
slowly rotating solar system planets: objects like \bpic~b that spend
their early evolution embedded in a disk experience some amount of
angular momentum braking, while free-floating objects are more free to
spin up.  \obj~c likely did not form in a disk, so measuring its
rotation from variability or \vsini\ would allow a direct test of
this idea.

As directly imaged objects, \bpic~b and \obj~c provide a new
opportunity to test atmospheric compositions and angular momentum
evolution for a close-in planet and a very wide companion that share a
common mass and age and that formed from the same material.

\vfill

\acknowledgments

This work would not have been possible without the excellent,
long-running support of CFHT staff past and present, including
Lo\"{\i}c Albert, Todd Burdullis, Pascal Fouqu\'e, Nadine Manset,
Karun Thanjavur, Kanoa Withington, and especially all the queue
observers who collected our WIRCam data.
We are indebted to Nancy Chanover, Russet McMillan, and Ben Williams
of Apache Point Observatory for kindly allowing us to conduct
TripleSpec observations during facility engineering time, and Ted
Rudyk for assistance with observations.
We are grateful to the $W$-band consortium for enabling our SpeX
observations, including Lo\"{\i}c Albert, Po-Shih Chiang, Bhavana
Lalchand, and especially Jessy Jose for observing assistance.
We thank Aaron Rizzuto for helpful discussions.
It is also a pleasure to thank the Keck Observatory staff for
assistance with the LGS AO observing, including Joel Aycock, Carolyn
Jordan, Marc Kassis, Gary Punawai, and Hien Tran.
We thank the anonymous referee for a timely review.
T.J.D.\ acknowledges research support from Gemini Observatory.
M.C.L.\ acknowledges support from NSF grant AST-1518339.
B.A.B.\ acknowledges support from STFC grant ST/J001422/1.
K.M.K.\ is supported in part by NSF grant AST-1410174.
A.W.M.\ was supported through Hubble Fellowship grant 51364 awarded by
the Space Telescope Science Institute, which is operated by the
Association of Universities for Research in Astronomy, Inc., for NASA,
under contract NAS~5-26555.
E.S.\ appreciates support from NASA/Habitable Worlds grant NNX16AB62G.
Our research has employed 
NASA's Astrophysical Data System;
the SIMBAD database and VizieR catalogue access tool, CDS, Strasbourg,
France;
and James R.\ A.\ Davenport's IDL implementation of the cubehelix color
scheme \citep{2011BASI...39..289G}.
This publication makes use of data products from the Wide-field
Infrared Survey Explorer, which is a joint project of the University
of California, Los Angeles, and the Jet Propulsion
Laboratory/California Institute of Technology, funded by the National
Aeronautics and Space Administration.
This publication utilizes data acquired with SpeX at the IRTF, which
is operated by the University of Hawaii under contract NNH14CK55B with
the National Aeronautics and Space Administration.
Finally, the authors wish to recognize and acknowledge the very
significant cultural role and reverence that the summit of Maunakea
has always had within the indigenous Hawaiian community.  We are most
fortunate to have the opportunity to conduct observations from this
mountain.

{\it Facilities:} \facility{Keck:II (LGS AO, NIRC2)}; \facility{CFHT (WIRCam)}; \facility{IRTF (SpeX)}; \facility{ARC (TripleSpec)}

\clearpage

\clearpage

\clearpage

\clearpage

\clearpage

\clearpage

\clearpage

\clearpage

\clearpage

\clearpage

\clearpage

\clearpage

\clearpage

\begin{landscape}
\begin{deluxetable}{lccccccccccccccc}
\setlength{\tabcolsep}{0.050in}
\tabletypesize{\scriptsize}
\tablewidth{0pt}
\tablehead{
\colhead{} &
\colhead{} &
\colhead{}  &
\multicolumn{4}{c}{\objlong} &
\colhead{}  &
\multicolumn{4}{c}{\objlongb} &
\colhead{}  &
\colhead{} &
\colhead{} &
\colhead{} \\
\cline{4-7}
\cline{9-12}
\multicolumn{2}{c}{Observation Date} &
\colhead{}  &
\colhead{R.A.}    &
\colhead{Dec.}    &
\colhead{$\sigma_{\rm R.A.}$} &
\colhead{$\sigma_{\rm Dec.}$}  &
\colhead{}  &
\colhead{R.A.}    &
\colhead{Dec.}    &
\colhead{$\sigma_{\rm R.A.}$} &
\colhead{$\sigma_{\rm Dec.}$}  &
\colhead{}  &
\colhead{Mean}    &
\colhead{Seeing}     &
\colhead{$\Delta$\Kn}      \\
\colhead{(UT)}   &
\colhead{(MJD)}  &
\colhead{}  &
\colhead{(deg)}  &
\colhead{(deg)}  &
\colhead{(mas)}  &
\colhead{(mas)}  &
\colhead{}  &
\colhead{(deg)}  &
\colhead{(deg)}  &
\colhead{(mas)}  &
\colhead{(mas)}  &
\colhead{}  &
\colhead{Airmass}  &
\colhead{(arcsec)} &
\colhead{(mag)} }
\tablecaption{Integrated-light \Kn-band astrometry from CFHT/WIRCam \label{tbl:cfht}}
\startdata
2011~Aug~6  & 55779.6419 && 042.48532936 & $-05.95995963$ &  4.9 &  4.1 && 042.47695124 & $-05.96729310$ &  8.8 &  6.4 && 1.137 & 0.54 & $3.796\pm0.026$ \\
2011~Sep~11 & 55815.5725 && 042.48533240 & $-05.95996130$ &  6.7 &  4.3 && 042.47695026 & $-05.96729258$ & 12.1 &  5.2 && 1.111 & 0.58 & $3.748\pm0.051$ \\
2011~Oct~16 & 55850.4559 && 042.48533009 & $-05.95996196$ &  3.7 &  3.1 && 042.47695126 & $-05.96729288$ &  5.0 &  5.9 && 1.126 & 0.50 & $3.780\pm0.031$ \\
2012~Aug~12 & 56151.6387 && 042.48534333 & $-05.95996614$ &  2.0 &  4.2 && 042.47696454 & $-05.96729865$ &  4.6 &  6.0 && 1.118 & 0.60 & $3.851\pm0.022$ \\
2012~Oct~5  & 56205.5070 && 042.48534321 & $-05.95996834$ &  3.3 &  3.4 && 042.47696800 & $-05.96730269$ &  5.4 &  6.0 && 1.110 & 0.58 & $3.764\pm0.021$ \\
2013~Oct~14 & 56579.5098 && 042.48535363 & $-05.95997709$ &  2.4 &  3.1 && 042.47697513 & $-05.96731035$ &  4.0 &  4.0 && 1.123 & 0.48 & $3.742\pm0.030$ \\
2014~Jul~30 & 56868.6432 && 042.48536427 & $-05.95998217$ &  2.7 &  5.1 && 042.47698636 & $-05.96731379$ &  6.1 &  5.3 && 1.172 & 0.50 & $3.776\pm0.022$ \\
2014~Oct~3  & 56933.5614 && 042.48536358 & $-05.95998275$ &  6.1 &  7.2 && 042.47698706 & $-05.96731761$ &  7.8 &  6.3 && 1.157 & 0.53 & $3.795\pm0.030$ \\
2014~Oct~13 & 56943.4851 && 042.48536301 & $-05.95998551$ & 10.4 &  5.6 && 042.47699064 & $-05.96731977$ &  5.5 &  7.5 && 1.110 & 0.63 & $3.743\pm0.041$ \\
2014~Oct~16 & 56946.4721 && 042.48536565 & $-05.95998450$ &  5.6 &  4.8 && 042.47698737 & $-05.96731859$ &  7.3 &  4.6 && 1.112 & 0.51 & $3.779\pm0.056$ \\
2015~Jan~21 & 57043.2480 && 042.48535865 & $-05.95998794$ &  3.5 &  2.8 && 042.47698235 & $-05.96732283$ &  5.9 &  8.2 && 1.133 & 0.57 & $3.801\pm0.032$ \\
\enddata
\tablecomments{The quoted uncertainties correspond to relative, not
  absolute, astrometric errors.}
\end{deluxetable}
\end{landscape}
\clearpage

\begin{deluxetable}{lcc}
\tabletypesize{\footnotesize}
\tablewidth{0pt}
\tablehead{
\colhead{Parameter} &
\colhead{\objlong} &
\colhead{\objlongb} }
\tablecaption{Parallax and Proper Motion from CFHT/WIRCam Astrometry \label{tbl:plx}}
\startdata

RA at first epoch\tablenotemark{*} (deg)            &\phs42.4853267 &\phs42.4769477 \\
Dec at first epoch\tablenotemark{*} (deg)           & $-$05.9599598 & $-$05.9672920 \\
\cline{1-3}
Relative parallax $\pi_{\rm rel}$ (mas)                   &\phs$19.2\pm2.1$ &\phs$18.8\pm3.5$ \\
Relative proper motion in RA (\masyr)  &\phs$39.4\pm1.0$ &\phs$42.5\pm1.6$ \\
Relative proper motion in Dec (\masyr) &   $-27.1\pm0.9$ &   $-28.7\pm1.4$ \\
\cline{1-3}
Absolute parallax $\pi_{\rm abs}$ (mas)                   &\phs$20.5\pm2.1$ &\phs$20.1\pm3.5$ \\
Absolute proper motion in RA (\masyr)  &\phs$42.9\pm2.0$ &\phs$46.0\pm2.3$ \\
Absolute proper motion in Dec (\masyr) &   $-30.2\pm1.8$ &   $-32.0\pm2.1$ \\
\cline{1-3}
$\chi^2$ (17 dof)  & \phs17.4 & \phs19.2 \\
$p(\chi^2)$        & \phs0.43 & \phs0.32 \\

\enddata
\tablenotetext{*}{First observation epoch: 55779.64~MJD, 2011~Aug~6~UT.}
\end{deluxetable}

\begin{deluxetable}{lcccc}
\tablewidth{0pt}
\tablehead{
\multicolumn{2}{c}{Observation Date} &
\colhead{Separation}  &
\colhead{PA}    &
\colhead{$\Delta{K}$} \\
\colhead{(UT)}     &
\colhead{(MJD)}    &
\colhead{(mas)}    &
\colhead{(deg)}    &
\colhead{(mag)}    }
\tablecaption{Keck LGS AO Astrometry of \obj{AB} \label{tbl:keck}}
\startdata
2012~Sep~7  & 56177.60 & $44.4\pm0.2$ & $233\fdg1\pm0\fdg3$ & $0.123\pm0.005$ \\
2013~Jan~17 & 56309.27 & $40.1\pm0.2$ & $237\fdg3\pm0\fdg5$ & $0.111\pm0.017$ \\
\enddata
\end{deluxetable}

\begin{deluxetable}{lcccc}
\tabletypesize{\footnotesize}
\tablewidth{0pt}
\tablehead{
\colhead{Property} &
\colhead{\obj{A}} &
\colhead{\obj{B}} & 
\colhead{\obj~c} &
\colhead{Notes} }
\tablecaption{Properties of the \obj\ System \label{tbl:prop}}
\startdata

Distance (pc)            & \multicolumn{3}{c}{$48.9^{+4.4}_{-5.4}$\phn} & 1 \\
$m-M$ (mag)              & \multicolumn{3}{c}{$3.44^{+0.21}_{-0.23}$} & 1\\
Age (Myr)                & \multicolumn{3}{c}{$22\pm6$} & 2 \\
                         
Spectral Type            & \multicolumn{2}{c}{M6~\vlg} & L2~\vlg & 3 \\
                         
$J$ (mag)                & \multicolumn{2}{c}{$11.885\pm0.027$\phn}& $16.64\pm0.06$\phn & 4 \\
$H$ (mag)                & \multicolumn{2}{c}{$11.410\pm0.026$\phn}& $15.61\pm0.06$\phn & 4 \\
$K$ (mag)                & $11.73\pm0.03$\phn & $11.85\pm0.03$\phn & $14.78\pm0.03$\phn & 4 \\
$J-K$ (mag)              & \multicolumn{2}{c}{$0.852\pm0.034$}     &  $1.86\pm0.05$     & 4 \\
$H-K$ (mag)              & \multicolumn{2}{c}{$0.376\pm0.033$}     &  $0.83\pm0.05$     & 4 \\

$W1$ (mag)               & \multicolumn{2}{c}{$10.844\pm0.023$\phn}&$14.125\pm0.034$\phn& 5 \\
$W2$ (mag)               & \multicolumn{2}{c}{$10.597\pm0.020$\phn}&$13.588\pm0.036$\phn& 5 \\
$W1-W2$ (mag)            & \multicolumn{2}{c}{$0.247\pm0.030$}     &  $0.54\pm0.05$     & 5 \\

$m_{\rm bol}$ (mag)      & $14.65\pm0.05$\phn & $14.77\pm0.05$\phn & $18.18\pm0.06$ & 6 \\
$\log(\Lbol)$ $[\Lsun]$  & $-2.59\pm0.09$\phs & $-2.64\pm0.09$\phs & $-4.00\pm0.09$ & 6 \\
Mass (\Mjup)             & $48^{+13}_{-12}$\phn & $44^{+14}_{-11}$\phn & $11.6^{+1.3}_{-1.0}$\phn & 7 \\

\multicolumn{5}{c}{} \\
                    & \multicolumn{3}{c}{Relative properties of AB--c} &  \\
\cline{1-5}
Separation (AU)     & \multicolumn{3}{c}{       $1950\pm200$\phn        } & 8 \\
Separation (arcsec) & \multicolumn{3}{c}{$39\farcs959\pm0\farcs005$\phn } & 8 \\
PA (deg)            & \multicolumn{3}{c}{$228\fdg649\pm0\fdg013$\phn\phn} & 8 \\
$\Delta\Kn$ (mag)   & \multicolumn{3}{c}{     $3.780\pm0.032$           } & 8 \\
\cline{1-5}

\multicolumn{5}{c}{} \\
                    & \multicolumn{2}{c}{Relative properties of A--B} & & \\
\cline{1-5}
Separation (AU)     & \multicolumn{2}{c}{$       2.17\pm0.22       $} & \nodata & 9  \\
Separation (arcsec) & \multicolumn{2}{c}{$0\farcs0444\pm0\farcs0002$} & \nodata & 9 \\
PA (deg)            & \multicolumn{2}{c}{$233\fdg1\pm0\fdg3$\phn\phn} & \nodata & 9 \\
$\Delta{K}$ (mag)   & \multicolumn{2}{c}{   $0.123\pm0.005$         } & \nodata & 9 \\

\enddata
\tablecomments{(1)~Computed directly from our measured parallax;
  (2)~lithium-depletion boundary age from \citet{2017AJ....154...69S};
  (3)~infrared types on the \citet{2013ApJ...772...79A} system;
  (4)~MKO photometry synthesized from SpeX spectra, where the
  integrated-light spectrum of \obj{AB} was flux calibrated using its
  2MASS photometry and \obj~c was flux calibrated from our CFHT/WIRCam
  \Kn-band photometry; (5)~AllWISE photometry
  \citep{2014yCat.2328....0C}; (6)~for \obj{AB} we used its
  integrated-light \mbol, observed $K$-band flux ratio, and assumed
  that the difference in $K$-band bolometric corrections for A and B
  is negligible; (7)~estimated from \citet{2015A&A...577A..42B} models
  for \obj{AB} and \citet{2008ApJ...689.1327S} hybrid models for
  \obj~c; (8)~from CFHT/WIRCam imaging; (9)~Keck/NIRC2 masking
  detection at discovery epoch 2012~Sep~7~UT.}
\end{deluxetable}
\clearpage

\begin{deluxetable}{lcccccc}
\tabletypesize{\tiny}
\tablewidth{0pt}
\tablehead{
\colhead{Name}         &
\colhead{Spectral Type} &
\colhead{$\pi$}    &
\colhead{\mbol}    &
\colhead{$\log(\Lbol/\Lsun)$}    &
\colhead{Mass}     &
\colhead{Ref.}     \\
\colhead{}     &
\colhead{}     &
\colhead{(mas)}    &
\colhead{(mag)}    &
\colhead{(dex)}    &
\colhead{(\Mjup)}    &
\colhead{}    }
\tablecaption{Late-type Members of the \bpic\ Moving Group \label{tbl:bpic}}
\startdata

PZ Tel B                                & M7                  &  $19.4\pm1.0$  & $14.44\pm0.15$ & $-2.45\pm0.07$ & $61\pm12$                & F15, M16, v07 \\
2MASSI~J0335020+234235                  & M7~\vlg             &  $21.8\pm1.8$  & $14.29\pm0.05$ & $-2.50\pm0.08$ & $56^{+13}_{-12}$         & AL13, D18, L16, S17 \\
HR~7329B                                & M7.5                & $20.74\pm0.21$ & $14.60\pm0.16$ & $-2.58\pm0.07$ & $49^{+12}_{-10}$         & L00, v07, F15 \\
\obj{A}                                 & M6~\vlg             &  $20.5\pm2.1$  & $14.65\pm0.05$ & $-2.59\pm0.09$ & $48^{+13}_{-12}$         & D18 \\
\obj{B}                                 & M6~\vlg             &  $20.5\pm2.1$  & $14.77\pm0.05$ & $-2.64\pm0.09$ & $44^{+14}_{-11}$         & D18 \\
SDSS~J044337.60+000205.2                & L0~\vlg             &  $47.3\pm1.0$  & $14.43\pm0.06$ & $-3.23\pm0.03$ & $20^{+4}_{-5}$\phn      & AL13, D18, L16, RB09 \\
\bpic~b                                 & L2\tablenotemark{a} & $51.44\pm0.12$ & $15.63\pm0.08$ & $-3.78\pm0.03$ & $13.0^{+0.4}_{-0.3}$\phn & D18, v07, M15 \\
\obj~c                                  & L2~\vlg             &  $20.5\pm2.1$  & $18.18\pm0.06$ & $-4.00\pm0.09$ & $11.6^{+1.3}_{-1.0}$\phn & D18 \\
PSO~J318.5338$-$22.8603                 & L7~\vlg             &  $45.1\pm1.7$  & $17.85\pm0.12$ & $-4.55\pm0.06$ &  $6.5^{+1.2}_{-0.8}$     & D18, L13, L16, A16 \\
51~Eri~b                                & T$6.5\pm1.5$        & $33.98\pm0.34$ &  $21.8\pm0.4$  & $-5.87\pm0.15$ & 2--12\tablenotemark{b}   & R17, v07 \\ 
\cline{1-7}
\multicolumn{7}{c}{} \\
\multicolumn{7}{c}{Possible Members ($\pi$ or RV unavailable, or ambiguous membership)} \\
\cline{1-7}
2MASS~J02241739+2031513                 & M6~\intg            &   \nodata      &  \nodata       & \nodata        & \nodata                  & S17 \\
2MASS~J03363144$-$2619578               & M6~\vlg             &   \nodata      &  \nodata       & \nodata        & \nodata                  & S17 \\
2MASS~J03370343$-$3042318               & M6~\fldg            &   \nodata      &  \nodata       & \nodata        & \nodata                  & S17 \\
2MASS~J19082195$-$1603249               & M6~\vlg             &   \nodata      &  \nodata       & \nodata        & \nodata                  & S17 \\
2MASS~J23355015$-$3401477               & M6~\vlg             &   \nodata      &  \nodata       & \nodata        & \nodata                  & S17 \\
2MASS~J22334687$-$2950101               & M7~\vlg             &   \nodata      &  \nodata       & \nodata        & \nodata                  & S17 \\
2MASS~J23010610+4002360                 & M7~\vlg             &   \nodata      &  \nodata       & \nodata        & \nodata                  & S17 \\
DENIS~J004135.3$-$562112AB              & M7.5~\vlg           &   \nodata      &  \nodata       & \nodata        & \nodata                  & G15 \\
2MASS~J03550477$-$1032415               & M8~\intg            &   \nodata      &  \nodata       & \nodata        & \nodata                  & S17 \\
2MASS~J19355595$-$2846343               & M9~\vlg             &  $14.2\pm1.2$  &  $15.9\pm0.2$  & $-2.76\pm0.11$ & $35^{+7}_{-15}$\phn      & AL13, D18, L16 \\
2MASS~J20004841$-$7523070               & M9~\vlg             &   \nodata      &  \nodata       & \nodata        & \nodata                  & G15 \\
2MASS~J00464841+0715177\tablenotemark{c}& L0~\vlg             &   \nodata      &  \nodata       & \nodata        & \nodata                  & G15, F16 \\
2MASS~J20135152$-$2806020               & L0~\vlg             &  $21.0\pm1.3$  & $16.18\pm0.05$ & $-3.22\pm0.06$ & $20^{+4}_{-6}$\phn       & AL13, D18, L16 \\
EROS-MP~J0032$-$4405\tablenotemark{d}   & L0~\intg            & \nodata        &  \nodata       & \nodata        & \nodata                  & AL13, G14, G15b \\
2MASSW~J2208136+292121\tablenotemark{e} & L3~\vlg             &  $25.1\pm1.6$  & $17.41\pm0.08$ & $-3.87\pm0.07$ & $12.6^{+0.7}_{-0.5}$\phn & AL13, D18, L16, V17 \\
\cline{1-7}

\multicolumn{7}{c}{} \\
\multicolumn{7}{c}{Candidates (proper motion only)} \\
\cline{1-7}
2MASS~J20334670$-$3733443               & M6~\intg            & \nodata        &  \nodata       & \nodata        & \nodata                  & G15 \\
2MASS~J01294256$-$0823580               & M7~\vlg             & \nodata        &  \nodata       & \nodata        & \nodata                  & G15 \\
2MASS~J02501167$-$0151295               & M7~\vlg             & \nodata        &  \nodata       & \nodata        & \nodata                  & G15 \\
2MASS~J05120636$-$2949540               & L5~\intg            & \nodata        &  \nodata       & \nodata        & \nodata                  & G15 \\
2MASS~J23542220$-$0811289               & L5~\vlg             & \nodata        &  \nodata       & \nodata        & \nodata                  & Sc17 \\
2MASS~J00440332+0228112                 & L7~\vlg             & \nodata        &  \nodata       & \nodata        & \nodata                  & Sc17 \\

\enddata

\tablenotetext{a}{Spectral resolution insufficient for gravity
  classification.}

\tablenotetext{b}{The luminosity of 51~Eri~b is low enough to be
  consistent with both hot-start and cold-start models, so its mass is
  correspondingly very uncertain.}

\tablenotetext{c}{\citet{2015ApJS..219...33G} reported this as a
  \bpic\ moving group candidate, but \citet{2016ApJS..225...10F}
  classify it as an ambiguous member after measuring an RV.}

\tablenotetext{d}{There are two published parallaxes for
  EROS-MP~J0032$-$4405. The value of $38.4\pm4.8$\,mas from
  \citet{2012ApJ...752...56F} used to determine \bpic\ membership by
  \citet{2014ApJ...783..121G, 2015ApJS..219...33G} is 1.8$\times$
  larger than the value of $21.6\pm7.2$\,mas from
  \citet{2013AJ....146..161M} used by \citet{2016ApJS..225...10F} to
  determine high-likelihood AB~Dor membership.}

\tablenotetext{e}{Our kinematic analysis indicates likely membership
  for \objtwo\ based on proper motion, parallax, and RV. But this is
  discordant with the results of BANYAN~$\Sigma$, so we consider the
  membership status ambiguous.}

\tablecomments{This table does not include nine new candidates
    identified by \citet{2018arXiv180511715G} using parallaxes and
    proper motions from the \Gaia~DR2 catalog because the objects have
    not been spectroscopically confirmed (photometrically estimated
    spectral types of M6--L3).}

\tablerefs{(A16)~\citet{2016ApJ...819..133A},
  (AL13)~\citet{2013ApJ...772...79A}, (D18)~this work,
  (F15)~\citet{2015ApJ...810..158F},
  (F16)~\citet{2016ApJS..225...10F},
  (G14)~\citet{2014ApJ...783..121G},
  (G15)~\citet{2015ApJS..219...33G},
  (G15b)~\citet{2015ApJ...798...73G},
  (L00)~\citet{2000ApJ...541..390L},
  (L13)~\citet{2013ApJ...777L..20L},
  (L16)~\citet{2016ApJ...833...96L},
  (M15)~\citet{2015ApJ...815..108M},
  (M16)~\citet{2016A&A...587A..56M},
  (R17)~\citet{2017AJ....154...10R},
  (RB09)~\citet{2009ApJ...705.1416R},
  (S17)~\citet{2017AJ....154...69S},
  (Sc17)~\citet{2017AJ....153..196S},
  (v07)~\citet{2007hnrr.book.....V},
  (V17)~\citet{2017ApJ...842...78V}.}

\end{deluxetable}

\end{document}